\documentclass[%
 aps,
 mph,%
 amsmath,amssymb,floatfix
 reprint,%
]{revtex4-2}
\usepackage{graphicx}
\usepackage{dcolumn}
\usepackage{bm}
\usepackage{multirow}
\usepackage[mathlines]{lineno}
\usepackage{longtable}
\usepackage{lineno}
\usepackage[T1]{fontenc}

\begin{document}

\title{Search for a viable nucleus-nucleus potential for heavy-ion nuclear reactions}
\author{T. Nandi$^{1\#}$, D.K. Swami$^1$, P.S. Damodara Gupta$^{2,3}$, Yash Kumar$^4$, S. Chakraborty$^1$, and H.C. Manjunatha$^{5*}$ }

\affiliation{\it $^{1*}$1003 Regal, Mapsko Royal Ville, Sector 82, Gurugram-122004, Haryana, India.}
\affiliation {\it $^2$Rajah Serfoji Govt. college affiliated to Bharathidasan University, Thiruchirappalli-620 024}
\affiliation {\it $^3$Department of Physics, Government First Grade College, Kolar-563101 Karnataka, India}
\affiliation {\it $^4$ Dipartimento di Fisica "Galileo Galilei", Università di Padova, I-35131 Padova, Italy}
\affiliation {\it $^5$ Department of Physics, Government college for women, Kolar-563101, Karnataka, India}
\thanks {for correspondence:\hspace{0.1cm} nanditapan@gmail.com\\Superannuated from Inter-University Accelerator Centre, Aruna Asaf Ali Marg, New Delhi-110067, India.}
\date{\today}

\begin{abstract}
We have constructed an empirical formulae for the fusion and interaction barriers using experimental values available till date. The fusion barriers so obtained have been compared with different model predictions based on the proximity, Woods-Saxon and double folding potentials along with several empirical formulas, time dependent Hartree-Fock theories, and the experimental results. The comparison allows us to find the best model, which is nothing but the present empirical formula only.  Most remarkably, the fusion barrier and radius show excellent consonance with the experimental findings for the reactions meant for synthesis of the superheavy elements also. Furthermore, it is seen that substitution of the predicted fusion barrier and radius in classic Wong formula [C. Wong, Phys. Rev. Lett. {31}, 766 (1973)] for the total fusion cross sections satisfies very well with the experiments. Similarly, current interaction barrier predictions have also been compared well with a few experimental results available and Bass potential model meant for the interaction barrier predictions. Importantly, the present formulae for the fusion as well as interaction barrier  will have practical implications in carrying out the physics research near the Coulomb barrier energies. Furthermore, present fusion barrier and radius provide us a good nucleus-nucleus potential useful for numerous theoretical applications.
\end{abstract}

\maketitle

\section{Introduction}\label{introduction}

\indent The basic characteristics of nuclear reactions are usually described by an interaction consisting of a repulsive Coulomb potential and a short range attractive nuclear potential. The resultant potential can be expressed as a function of the distance between the centres-of-mass of the target and projectile nuclei. When a projectile ion approaches the target nucleus it experiences a maximum force at a certain distance where the repulsive and attractive forces cannot balance each other, the repulsive Coulomb force is always higher. The projectile ion needs to overcome the barrier for coming closer to the target nucleus. This barrier is referred to the fusion barrier $(B_{fu})$, which is a basic parameter in describing the nuclear fusion reactions and is simply defined as the maximum of the total potential without the centrifugal term.  The kinetic energy of the projectile ion must be adequate to surmount this barrier in order to enter a pocket at the adjacent to the barrier at the shorter distances, where the nuclei undergo the nuclear fusion processes. Furthermore, the fusion barrier height and its width play a crucial influence on the tunneling process during the sub-barrier fusion. The situation is further complicated by the presence of multiple barriers. The fusion barrier is determined by the excitation function measurement of the nuclear fusion experiments \cite{back2014recent}, whereas it is estimated by many theoretical models such as the Bass potential model \cite{bass1974fusion,bass1973threshold}, proximity potential model \cite{blocki1977proximity}, double folding model \cite{satchler1979folding}, and semi-empirical models such as Christensen and Winther (CW) model \cite{christensen1976evidence}, Broglia and Winther (BW) model \cite{reisdorf1994heavy}, Aage Winther (AW) model \cite{winther1995dissipation}, Denisov potential (DP) model \cite{dutt2010systematic}, Siwek-Wilczyńska and Wilczyński (SW) model \cite{siwek2004empirical}, Skyrme energy density function (SEDF) model \cite{wang2006applications} and the Sao Paulo optical potential (SPP)\cite{freitas2016woods}.
\indent In contrast, the quasi-elastic (QEL) processes, involving smaller energy transfer  due to single nucleons or clusters such as alpha particles than the fusion reactions, excite only nuclear levels in either one of the participating nuclei or in both as soon as the two bodies approach within the range of the nuclear forces. The position where the resultant of the Coulomb and nuclear forces is still repulsive and additional energy is required to get the two nuclei interacting over their mutual potential barrier. This barrier is somewhat smaller than the fusion barrier and is known as the interaction barrier $(B_{int})$, which is measured by the excitation function studies of QEL scattering experiment \cite{mitsuoka2007barrier}. Obviously, both the barriers are Coulomb barriers, but they are different from each other. One is characterized by the fusion reaction when the two nuclei fuse to form a compound nucleus  and the other by the QEL scattering when the two nuclei  enter barely to the strong force regime \cite{bass1974fusion,bass1973threshold} keeping their identities almost intact. However, many-a-time the distinction is overlooked, for example,  \cite{mitsuoka2007barrier,dutt2010comparison}, though the concept was introduced in seventies \cite{bass1974fusion,bass1973threshold}.  Worth noting here that it is only the Bass who has segregated the appearance of the Coulomb barrier in above mentioned, two different ways: one is the Bass interaction model and the other is the Bass fusion model \cite{bass1974fusion,bass1973threshold}.\\
\indent Recently Sharma and Nandi \cite{sharma2017shakeoff} demonstrated the coexistence of the atomic and nuclear phenomenon on the elastically scattered projectile ions while approaching the Coulomb barrier. Here the projectile ion x-ray energies were measured as a function of ion beam energies for three systems $^{12}$C($^{56}$Fe,$^{56}$Fe), $^{12}$C($^{58}$Ni,$^{58}$Ni) and $^{12}$C($^{63}$Cu,$^{63}$Cu) and observed unusual resonance like structures as the beam energy approaching the fusion barrier energy, according to the Bass model \cite{bass1974fusion,bass1973threshold}. However, the resonance would have occurred near the interaction barrier as the technique involved only the elastic phenomenon and thus resembling the quasi-elastic (QEL) scattering experiment \cite{mitsuoka2007barrier}. It implies the resonance should have appeared adjacent to the interaction barrier. To resolve this anomaly, we planned to examine both the fusion and interaction barriers in a greater detail. Note that present attempt is not to make a concrete theoretical model to describe various possible steps of a reaction leading to the final products, but to find the best model available till date so that the above anomaly be resolved. Besides the existing models, we have used the experiments found in the literature to construct an empirical formula for estimating the fusion barriers from the fusion excitation function measurements alone and another for the interaction barriers from the QEL scattering experiments only.\\ 
\indent In the next step, we have compared the empirical model predictions for fusion barrier with various models based on proximity type of potentials such as Bass potential \cite{bass1974fusion,bass1973threshold} and Christensen and Winther (CW) \cite{christensen1976evidence} and Woods-Saxon type of potentials such as Broglia and Winther (BW) \cite{reisdorf1994heavy}, Aage Winther (AW) \cite{winther1995dissipation}, Siwek-Wilczyńska and Wilczyński (SW) \cite{siwek2004empirical}, SEDF \cite{wang2006applications} models, and the Sao Paulo optical potential (SPP) \cite{freitas2016woods}.
Further, the present interaction barrier formula has been compared with the Bass interaction model \cite{bass1974fusion,bass1973threshold}. It is seen that this work will be useful in various applications \cite{back2014recent}, for example, prediction of $B_{fu}$ or $V_B$ for the formation of the superheavy elements \cite{banerjee2015fusion} and that of both $B_{fu}$ and $B_{int}$ for the significant physics research near the Coulomb barriers \cite{sharma2017shakeoff}. 
\section{Determination of the barriers}
\indent Mean fusion barrier height may be obtained from the Gaussian fit of the barrier distribution plot, which is defined as the second derivative of the energy-weighted cross section $\frac{d^2(\sigma E)}{dE^2}$ (point difference) versus beam energy in the center-of-mass frame \cite{bass1977nucleus,rowley1991distribution}. In many articles only the excitation function is reported, we have converted it into a barrier distribution plot to obtain the mean fusion barrier.\\
\indent Similarly, the interaction barrier can be obtained from the QEL excitation function studies. The QEL scattering is affected by the sum of elastic, inelastic, and transfer processes, which is measured at backward angles of nearly $180^0$, where the head-on collisions are dominant. The barrier distribution is obtained by taking the first derivative, with respect to the beam energy, of the QEL cross-section relative to the Rutherford cross section, that is, $\frac{-d}{dE} (\frac{d\sigma_{QEL}}{d\sigma_R})$ \cite{andres1988effect}. This method has been examined in several intermediate-mass systems \cite{timmers1995probing,hagino2004large}. One can notice that the QEL barrier distribution behaves similarly to the fusion barrier distribution, although the former is less sensitive to the nuclear structure effects. \\
\section{General background}
Theoretically, the total nucleus-nucleus interaction potential $V_T(r)$ between the projectile and  target nuclei, in general, is written as a function of the distance $r$ between the two nuclei  
\begin{equation}
V_T(r)=V_N (r) + \frac{l(l+1)\hbar^2}{2\mu r^2} + V_c (r).
\label{eq1}
\end{equation}
where the first term $V_N(r)$ is the model dependent nuclear potential, the second term is centrifugal potential so that $\mu = \frac{A_p A_t}{A_p+A_t}$ is the reduced mass of the projectile mass $ A_p$ and the target nuclei mass $A_t$ in MeV/$c^2$ units and $l$ represents the angular momentum of the two body system. When we consider the fusion barrier of the system, $l$ is set to zero which means the centrifugal or the second term is zero. The third term is the Coulomb potential as given by \cite{birkelund1979heavy} 
\begin{equation}
 V_c(r) = \frac{Z_1 Z_2 e^2}{4\pi\epsilon_0}
  \begin{cases}
    \frac{1}{r}       & \quad \text{for } r \geq R_B\\
    \frac{1}{2R_B} \Big[3-\Big(\frac{R_B}{r}\Big)^2\Big]  & \quad \text{for}\; r<R_B.
  \end{cases}
   \label{eq2}
\end{equation}
Here the fusion barrier radius $R_B$ = $R_c$($A_p^\frac{1}{3}$ + $A_t^\frac{1}{3}$), $R_c$ depends on the system as discussed below. Putting the first term of equation (\ref{eq1}) from any particular model, one can solve the fusion barrier radius $R_B$ by using the conditions
\begin{equation}
V_T(r=R_B) = B_{fu}\;,\;\; \frac{dV_T(r)}{dr}\bigg|_{r=R_B} = 0\;\;~ and ~\;\; \frac{d^2V_T(r)}{dr^2}\bigg|_{r=R_B} \leq 0.
 \label{eq3}
\end{equation}
Also,  $V_T(r=R_{int})$ = $B_{int}$, where $R_{int}$ is the interaction barrier radius.  Of course, $V_N(r)$ in equation (\ref{eq1}) may need to be replaced by another appropriate potential to obtain the interaction barrier; for example, the Bass potential model \cite{bass1973threshold} uses different potential forms to estimate the fusion and interaction barriers.\\
 \section{Present empirical formula}
\begin{figure}
    \centering
    \includegraphics{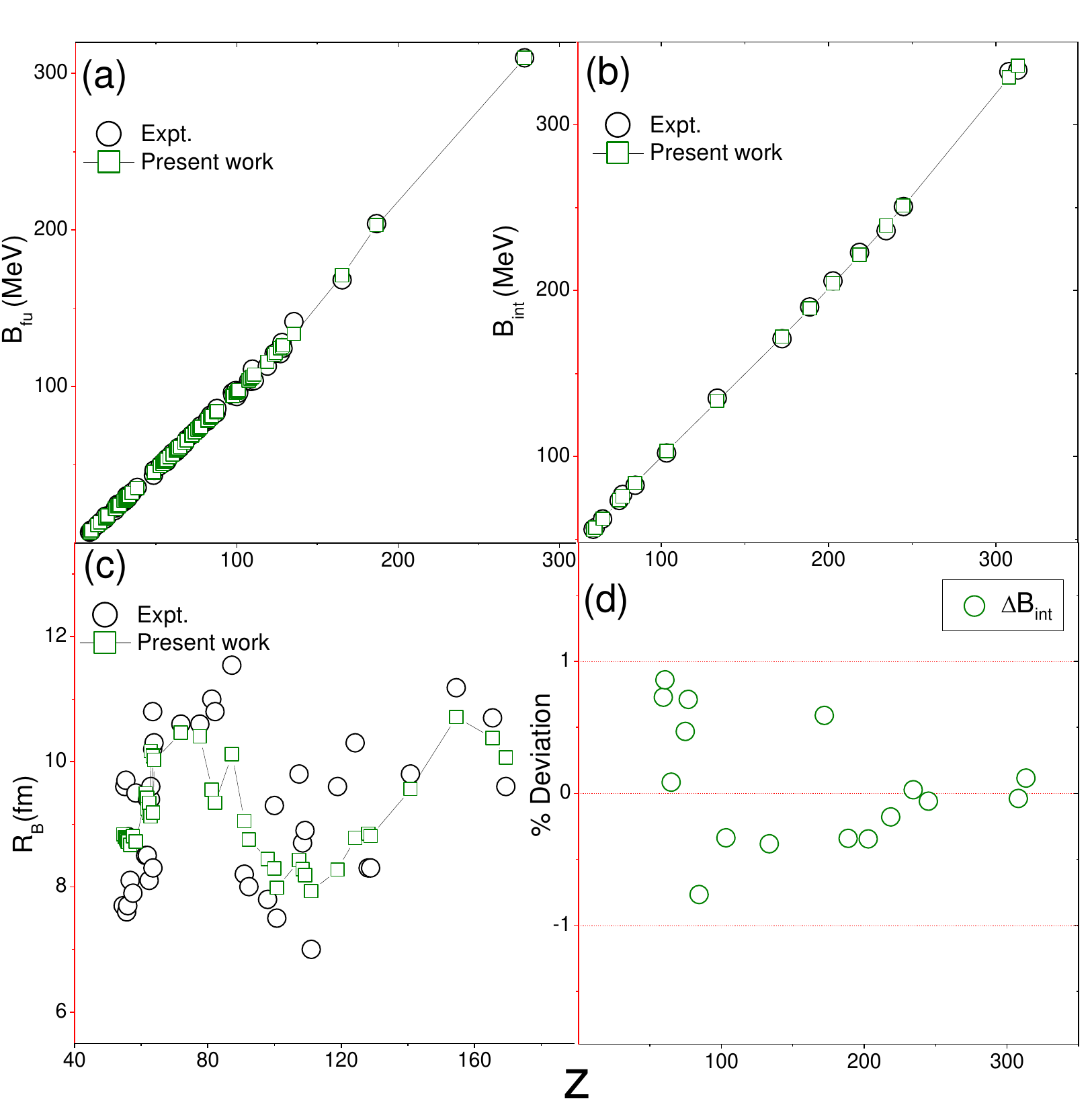}
    \caption{Fusion barrier, interaction barrier, fusion barrier radius and percentage of deviation plots: (a) the fusion barrier B$_{fu}$ (MeV), as obtained from the fusion excitation function experiments, (b) the interaction barrier B$_{int}$ (MeV) from the quasi-elastic scattering experiments, (c) the fusion barrier radius R$_{B}$ (fm), (d) percentage of deviation of present value from the experimental B$_{int}$ are plotted against the dimensionless parameter z. The percentage of deviation for B$_{fu}$ and R$_{B}$ are shown in figures 2 and 3.  }
    \label{f1}
\end{figure}

\indent According to the definition for the Coulomb potential given above, the shape of the nuclear potential discussed below and representation of nuclear distances, for example, one given above for the barrier radius, $B_{fu}$ and $B_{int}$ may be written as a function of $Z_p$, $Z_t$, $A_p$ and $A_t$. Hence, the experimentally obtained $B_{fu}$ from fusion excitation function measurements and $B_{int}$ from QEL measurements can be plotted against the Coulomb interaction parameter $z = \frac{Z_p Z_t}{(A_p^\frac{1}{3}+A_t^\frac{1}{3} )}$, as shown in Fig.\ref{f1}(a) and Fig.\ref{f1}(b), respectively. The fusion experiments used for Fig. \ref{f1}(a) and the QEL experiments used for Fig. \ref{f1}(b) are given in Table \ref{t1} and \ref{t2}, respectively. Fusion data are available for $8\leq z \leq286$, whereas QEL data for $59\leq z \leq313$. We can notice that both $B_{fu}$ vs $z$ and $B_{int}$ vs $z$ are nonlinear. The whole range of $B_{fu}$-data has been fitted by two non-linear functions of a sixth-degree polynomial to obtain the reduced chi-square nearly equal to one. Whereas, the full range of $B_{int}$-data is fitted with a single non-linear function.\\
\indent The polynomial function that fits the  $B_{fu}$ vs $z$ data extremely well is as follows
\begin{align}
  B_{fu}^{(1)}=&-1.2725+0.9106z+5.6932\times 10^{-4}z^2+2.335\times 10^{-5}z^3\label{eq4}\\\nonumber& -4.4975\times 10^{-7}z^4+2.7836\times 10^{-9}z^5-5.2482\times 10^{-12}z^6\; \mbox{ for}\; 8 \leq z  \leq 128
\end{align}
and
\begin{align}
 B_{fu}^{(2)}=&-34488.7618+1100.6666z-14.4066z^2+9.9275\times10^{-2}z^3-3.7959\times10^{-4}z^4\label{eq5}\\\nonumber&+7.6357\times10^{-7}z^5-6.3136\times10^{-10}z^6 \;\mbox{for}\;128 \leq z  \leq 286  
\end{align}
     and the other function that fits the $B_{int}$ vs $z$ data is given by
\begin{align}
B_{int}=0.09295z^{3/2}-2.1601z+35.5879z^{1/2}-132.8943\;\mbox{ for}\; 59 \leq z \leq 313
\label{eq6}
\end{align}
    The $B_{fu}$ can be predicted using equation (\ref{eq4}) for any system within $8 \leq z  \leq 128$  and equation (\ref{eq5})  for any system within $128 \leq z  \leq 286$.  Such an empirical formula was constructed in the past \cite{swiatecki2005fusion} as follows
\begin{align}
B_{WJS}=0.85247z + 0.001361z^2 - 0.00000223z^3 MeV.
\label{eqSJW}
\end{align}
It was a polynomial function of z of degree 3, where as our empirical formula contains up to a degree of 6.

\indent To obtain $R_B$, we follow a method involving the reduced fusion barrier position S$_B$ (the separation between the half-density surfaces of the two nuclei) \cite{ manjunatha2018pocket}
\begin{align}
    S_B=R_B^{exp}-C_1-C_2
\end{align}
Where the half-density radius of the matter
    distribution $C_i=\left(R_i-\frac{b^2}{R_i}\right)$ (for i=1,2) \cite{myers2000nucleus}, the sharp radius  $R_i=1.233A_i^{1/3}-0.98A_i^{-1/3}$ and  the measure of the diffuseness of the nuclear surface b=0.99 fm. $S_B$ is fitted with 4$^{th}$ order polynomial as a function of z as
\begin{align}
    S_B=-46.089+2.0478z-3.0962\times 10^{-2}z^2+1.9278\times 10^{-4}z^3-4.2492\times 10^{-7} z^4
\end{align}
and the R$_B$ is obtained from the reduced fusion barrier position
\begin{align}
    R_B=S_B+C_1+C_2 \label{eq7}
\end{align}
\section{Nuclear potential models in literature}
\indent In this section we present different nucleus-nucleus potentials $V_N(r)$, which can be used for obtaining the fusion and interaction to compare with the available measured values.
\subsection{Bass potential model}
\indent Bass potential model \cite{bass1974fusion,bass1973threshold} suggests that the Coulomb barrier for quasi-elastic surface reaction is in general different from the Coulomb barrier for fusion. The former results from the quasi-elastic processes,  where not much mass or energy transfer takes place, whereas maximum mass and energy can transfer in the latter. Further, the quasi-elastic processes become significant as the projectile and target nuclei approach to the range of nuclear force, where the resultant of the Coulomb and nuclear forces is still repulsive. Thus additional energy is required to get the nuclei within the resultant attractive force, where the fusion can occur. According to the Bass interaction model, the barriers responsible in the quasi-elastic reactions is called the interaction barrier $B_{int}$ and it can be determined from the elastic scattering experiments \cite{mitsuoka2007barrier}. The other barrier is significant in fusion reactions and is defined as the fusion barrier $B_{fu}$. The latter is equal to the height of the potential barrier for zero angular momentum.\\
\indent The total effective Bass potential consists of a Coulomb, nuclear and centrifugal terms
  \begin{equation}
 V_l (r)= \frac{Z_p Z_t e^2}{4\pi\epsilon_0 r} + \frac{\hbar^2 l^2}{2\mu r^2} - a_s  A_p^\frac{1}{3}  A_t^\frac{1}{3} \frac{d}{R_{pt}}  {e^{-(\frac{r-R_{pt}}{d})}}
 \label{eq12}
  \end{equation}
  Where $d$ is the range of nuclear interaction. The influence of fragment (projectile or target nuclei) properties on this potential can be expressed in terms of the dimensionless parameters 
  \begin{equation}
  x = \frac{e^2}{r_0 a_s4\pi\epsilon_0}   \frac{Z_p Z_t}{A_p^\frac{1}{3} A_t^\frac{1}{3} (A_p^\frac{1}{3} + A_t^\frac{1}{3} )}
     \label{eq13}
    \end{equation}
    and 
    \begin{equation}
    y= \frac{\hbar^2}{2m_0 r_0^2 a_s} \frac{A_p+A_t}{A_p^\frac{4}{3} A_t^\frac{4}{3} (A_p^\frac{1}{3} + A_t^\frac{1}{3})^{2}}.
     \label{eq14}
    \end{equation}
    Where $x$ is the ratio of the Coulomb force to the nuclear force and $yl^2$ is the ratio of the centrifugal force to the nuclear force at the point of contact i.e., $r= R_{pt} = r_0 (A_p^\frac{1}{3}+A_t^\frac{1}{3} )$, $r_0$=1.07 fm. Here,  $a_s$ =17.23 MeV is the surface constant as used in the liquid drop model of fission, $m_0$ the mass of a nucleon and other notations have usual significance.  Whereas $B_{fu}$ acts at $r$ = $R_{pt}$ + $d_{fu}$, $d_{fu}$ is the fusion distance. The $B_{int}$ is applicable for an interaction distance between the two surfaces $(d_{int})$ or the centre to centre distance, $R_{int}$ = $R_{pt}$ + $d_{int}$. The $d_{int}$ is always longer than the $d_{fu}$, which can be approximately obtained from the relation
     \begin{equation}
   \frac {d_{fu}}{d} \approx -\frac{lnx}{(1- \frac{2d}{R_{pt}} )}
         \label{eq15}
         \end{equation}
The $d_{fu}$ varies with the fragments in the nuclear interactions. The barriers $B_{fu}$ and $B_{int}$ can be obtained from \\
    \begin{equation}
 B_{fu} = \frac{Z_p Z_t e^2}{4\pi\epsilon_0 R_{pt}}  \Big\{\frac{R_{pt}}{R_{pt}+d_{fu}} - \frac{1}{x} \frac{d}{R_{pt}}   e^{\big(-\frac{d_{fu}}{d}\big)}\Big\}
     \label{eq16}
    \end{equation}
    \begin{equation}
  B_{int} =\frac{Z_p Z_t e^2}{R_{pt}+d_{int}}- 2.90 \frac{A_p^\frac{1}{3} A_t^\frac{1}{3}}{(A_p^\frac{1}{3}+A_t^\frac{1}{3})}
      \label{eq17}
      \end{equation}
      \begin{equation}
      d_{int}=2d = 2 \times 1.35=2.70 fm
           \label{eq18}
           \end{equation}
 Here, it is assumed in equation (\ref{eq18}) that $d$ is independent of the mass of the nuclei.
  \subsection{Christensen and Winther model}
 \indent  Christensen and Winther (CW) \cite{christensen1976evidence} use elastic scattering trajectory leading to the rainbow pattern that is strongly connected to low-lying target excitation. Hence, the phenomenon happens above the fusion barrier. They derived the nucleus-nucleus interaction potential to represent the nuclear fusion on the basis of semi classical arguments as given by 
  \begin{equation}
   V_N^{CW}(r) = - 50 {\overline{R}}{ e^{\big(-\frac{r-R_{pt}}{a}\big)}} MeV
   \label{eq19}
   \end{equation}
   where $R_{pt}$ = $R_p+R_t$,  $\overline{R}  = \dfrac{R_p R_t}{R_p+R_t}$ and $a$ is the diffuseness parameter ($a= 0.63 fm$). This form is similar to that of the Bass model \cite{bass1974fusion} with different sets of radius parameter.
    \begin{equation}
    R_i = 1.233 A_i^\frac{1}{3}-0.98 A_i^{-\frac{1}{3}} fm~(i=p,t)
     \label{eq20}
     \end{equation}
     Here, the radius of the fusion barrier has the form 
       \begin{equation}
       R_B = 1.07 (A_p^\frac{1}{3}+A_t^\frac{1}{3} )+2.72  fm
        \label{eq21}
        \end{equation}
        and  the total nucleus-nucleus potential for $l=0$ is 

        \begin{equation}
            U^{CW} (r) = \dfrac{Z_{p}Z_{t}e^{2}}{4\pi\epsilon_0 r} + V_N^{CW} (r)
             \label{eq22}
             \end{equation}
        and thus, the fusion barrier can be obtained from $U^{CW}(r=R_B$).
        
       \subsection{Broglia and Winther model}  
       Broglia and Winther (BW) \cite{reisdorf1994heavy} have refined the CW potential \cite{christensen1976evidence} in order to make it compatible with the value of the maximum nuclear force of the proximity potential \cite{blocki1977proximity}. This refined force is taken as the standard Woods-Saxon potential given by
        \begin{equation}
        V_N(r) = \dfrac{-V_{0}}{1+e^\frac{r-R_{pt}}{a}} MeV
         \label{eq23}
         \end{equation}
         with $V_0$ = $16\pi a\gamma\frac{R_{p}R_{t}}{R_{p}+R_{t}} , a=0.63 fm$ and  
         \begin{equation}
           R_{pt} = R_p+R_t+0.29 fm  
           \label{eq24}
         \end{equation}
         Here the nucleus radius $R_{i}$ is given by:
          \begin{equation}
          R_i = 1.233A_i^\frac{1}{3}-0.98A_i^\frac{-1}{3}  fm      (i=p,t)
          \label{eq25}
          \end{equation}
          The surface energy coefficient ($\gamma$) has the form
           \begin{equation}
           \gamma = \gamma_0 \bigg[1-k_s \bigg(\frac{N_{p}-Z_{p}}{A_{p}}\bigg)\bigg(\frac{N_{t}-Z_{t}}{A_{t}}\bigg)\bigg] MeV fm^2
            \label{eq26}
            \end{equation}
            Where $\gamma_0$ = $0.95 MeVfm^{-2}$  and $k_s = 1.8$.
            The total interaction potential of the two heavy ions is
            \begin{equation}
            U^{BW} (r) = \dfrac{Z_{p}Z_{t}e^{2}}{4\pi\epsilon_0 r} + V_N^{BW} (r)
             \label{eq27}
             \end{equation}
              and it displays a maximum, i.e., the fusion barrier and the barrier radius $(R_B )$ is the solution of the following equation
              \begin{equation}
             \dfrac{dU^{BW}(r)}{dr}|_{r=R_B} =-\dfrac{Z_{p}Z_{t}e^{2}}{4\pi\epsilon_0 R_{B}^{2}}+ \dfrac{V_{0}\hspace{2mm} e^\frac{R_{B}-R_{pt}}{a}}{{a\hspace{1mm} (1+e^{\frac{R_{B}-R_{pt}}{a}})}^{2}}=0
                        \label{eq28}
               \end{equation}
               and  $ U^{BW} (R_B )$ is the fusion barrier.
               
                \subsection{Aage Winther model}
               Aage Winther (AW) \cite{winther1995dissipation} adjusted slightly the parameters of the Broglia and Winther potential through an extensive comparison with the experimental data for heavy-ion elastic scattering. The resulting values of $a$ and $R_i$ are as follows:
               \begin{equation}
               a = \Bigg[\dfrac{1}{1.17(1+0.53(A_p^{-\frac{1}{3}}+A_t^{-\frac{1}{3}} )) }\Bigg]  \hspace {2mm}fm
               \label{eq29}
               \end{equation}
               and 
               \begin{equation}
               R_i = 1.20A_i^\frac{1}{3}-0.09 \hspace{2mm}fm           \hspace{9mm} (i=p,t)
                \label{eq30}
                \end{equation}
                and $R_{pt}$ of the BW model is written as $ R_{pt}$ = $R_p+R_t$ only.
                
                \subsection{Siwek-Wilczyńska and Wilczyński model}
                Siwek-Wilczyńska and Wilczyński (SW) \cite{siwek2004empirical} have used a large number of reactions to determine an effective nucleus-nucleus potential for reliable prediction of the fusion barriers for the systems that are studied. In this approach the nucleus-nucleus potential is taken also as the Woods-Saxon shape is given in equation (\ref{eq23}).
                  Where $R_{pt}$ = $ R_p+R_t$ and $R_i =  R_c A_i^\frac{1}{3}$ $(i=p,t)$, the radius parameter $R_c$ is constant, $a$ is the diffuseness parameter and $V_0$ is the depth of the potential. The $V_0$ is given by
                    \begin{equation}
                    V_0 = V_0^{'}+ S_{cn}.
                     \label{eq31}
                     \end{equation}
                     Where $S_{cn}$ is the shell correction energy \cite{moller1993nuclear} and 
                     \begin{center}
                     $V_0^{'} = (M_p+ M_t- M_{cn} ) c^2+ C_{cn}-C_p-C_t$
                     \end{center}
                     
                     \begin{equation}
                     = Q_{fu}+C_{cn}-C_p-C_t.
                     \label{eq32}
                     \end{equation}
                     Here $Q_{fu}$ is the ground state $Q$ value for fusion and $C_i$ are the intrinsic Coulomb energies \cite{siwek2004empirical} as given by
                      \begin{center}
                     $C_{cn} - C_p - C_t = C_0$
                     \end{center}
                      \begin{equation}
                     C_0 = 0.7054\Bigg[\dfrac{(Z_p+Z_t )^2}{(A_p+A_t )^\frac{1}{3}} - \dfrac{Z_p^2}{A_p^\frac{1}{3}}-\dfrac{Z_t^2}{A_t^\frac{1}{3}}\Bigg] \hspace{4mm} MeV.
                     \label{eq33}
                     \end{equation}
                     The Coulomb energy constant is taken from the standard liquid-drop-model fit to nuclear masses \cite{Myers1967}. Now, the equation (\ref{eq31}) can be written as
                     \begin{equation}
                     V_0= Q_{fu}+C_0+ S_{cn}.
                      \label{eq34}
                      \end{equation}
                      For determination of the fusion barrier, one considers the nucleus-nucleus potential in the region $R > R_{pt}$ as
                      \begin{equation}
                      V(r) = V_N (r)+ \dfrac{Z_p Z_t e^2}{4\pi\epsilon_0 r}.
                       \label{eq35}
                       \end{equation}
                        For the region $r < R_p+R_t, e^{\big[\frac{r-R_{pt}}{a}\big] }\longrightarrow0$, the nucleus-nucleus potential takes the form
                        \begin{equation}
                        V(r) = C_0-V_0= -Q_{fu}-S_{cn}.
                         \label{eq36}
                         \end{equation}
                         Equation (\ref{eq35}) gives thus the fusion barrier. It has only two free parameters $R_c$ and $a$ as $V_0$ is known from equation (\ref{eq34}). These parameters are obtained by fitting the barriers from equation (\ref{eq35}). The experimental $B_{fu}$ values can be obtained where the measured fusion excitation functions are fitted with the following expression
                           \begin{equation}
                           \sigma_{fus} = \pi r_{\sigma}^{2} \dfrac{\omega}{E\surd2\pi} [X\surd\pi (1+erf(X))+exp(-X^{2}].
                            \label{eq37}
                            \end{equation}
                            Where $X = \dfrac{E-B_{fu}}{\sqrt{2} \omega}$ and  erf(X) is the Gaussian error integral of the argument X as
                            \begin{equation}
                        erf(X) = \dfrac{1}{\sqrt{\pi}}{ \int_{0}^{X} e^{-t^{2}}dt}.
                             \label{eq38}
                             \end{equation}
                             The fitting gives three parameters the fusion barrier $B_{fu}$, the relative distance corresponding to the position of the approximate barrier $r_\sigma$ and the width of the barrier $w$.
                             However, the values of $R_c$ and $a$ depend on the Coulomb barrier parameter $z$, for example,\\
                             $R_c =1.25 fm$ and $a=0.481 fm$\hspace{5mm} for$ z<70$\\
                            $ R_c$ = $1.18 fm$ and $a = 0.675 fm$    \hspace{5mm}   for    $70<z<130$\\
$R_c=1.11 fm$ and $a=0.895 fm$ \hspace{5mm}for $z>130$

\subsection{Skyrme energy density function model}
Skyrme energy density function (SEDF) model has been introduced by Wang et al. \cite{wang2006applications,liu2006applications}, where the total binding energy of a nucleus is represented as the integral of the energy density function \cite{bartel2002nuclear} 
\begin{equation}
E = \int{Hdr}.
\label{eq39}
\end{equation}
Here energy-density function $H$ has three parts: kinetic energy, Coulomb and nuclear interactions and is generally defined as follows:
\begin{equation}
H(r)=\frac{{\hbar}^2}{2m}\left[ \tau_{p}\left( r\right) +\tau_{n}\left( r\right) \right]+H_{coulomb}\left( r\right) +H_{nuclear}\left( r\right) 
\label{eq40}
\end{equation}
Where $\tau_{p}$ and $\tau_{n}$ are the kinetic energy density for proton and neutron, respectively.
The interaction potential $V_{B}\left( R\right) $ is defined as
\begin{equation}
V_{B}\left( R\right)  = E_{tot}\left( R\right) - E_{p} - E_{t}
\label{eq41}
\end{equation}
where $E_{tot}\left( R\right) $ is the total energy of the interacting nuclear system, $E_{p}$ and $E_{t}$ are the energies of the projectile and target, respectively, at completely separated distance $R$. These energies can be calculated by the following relation
\begin{equation*}
 E_{tot}(R) =\int{H[ \rho_{1p}(r) +\rho_{2p}( r-R),\rho_{1n}(r) +\rho_{2n}( r-R)] dr}   
\end{equation*}
\begin{center}
$E_{p}\left( R\right) =\int{H\left[ \rho_{1p}\left( r\right) ,\rho_{1n}\left( r\right) \right] dr}$
\end{center}
\begin{equation}
E_{t}\left( R\right) =\int{H\left[ \rho_{2p}\left( r\right) ,\rho_{2n}\left( r\right) \right] dr}
\label{eq42}
\end{equation}
The densities of the neutron $\rho_{n}$ and proton $\rho_{p}$ for projectile and target can be described by the spherically symmetric Fermi function
\begin{equation}
\rho{\left( r\right)} =\dfrac{{\rho}_{0}}{1+{e^{\frac{\left( r-c\right)}{a}}}}
\label{eq43}
\end{equation}
where $\rho_{0}$, $c$ and $a$ are the parameters of the densities of the participating nuclei in the reactions, which are obtained by using the density-variation approach and minimizing the total energy of a single nucleus given by the SEDFM \cite{wang2006applications,liu2006applications}.

Using the Skyrme energy density formalism, Zanganeh et al. \cite{zanganeh2015analytical} have constructed a pocket formula for fusion barriers and positions in the range 8$\le$ z $\le$ 168 with respect to the charge and mass numbers of the interacting nuclei as follows:
\begin{equation}
    V_B^{Par}=-0.01[(Z_p Z_t)(A_p^\frac{1}{3}+A_t^\frac{1}{3})]+0.20(Z_p Z_t)+0.60
    \label{eq44}
\end{equation}
\begin{equation}
    R_B^{Par}=1.40[A_p^\frac{1}{3} + Z_t^\frac{1}{3}] - 0.07(Z_p Z_t)^{0.05} +1.40
    \label{eq45}
\end{equation}
We have made use of the equation (\ref{eq44},\ref{eq45}) for SEDFM predictions for various reactions as shown in Table \ref{t2} and Table \ref{t3}. Since, the SEDFM is based on the frozen density approximation, predicted values for each of the considered fusion systems are a bit higher than the corresponding experimental data. 
\subsection{Sao Paulo optical potential (SPP)} 
This model \cite{freitas2016woods} also takes a Woods-Saxon form for the nuclear potential as given in equation (\ref{eq23}). In the approximation of $\exp{(\frac{R_B-r}{a})}\gg 1$, the $R_B$ can be written as:
\begin{equation}
R_B = r + 0.65 \ln{x}
\label{eq48}
\end{equation}
where $x = 27.1 \times \frac{A_p^\frac{1}{3}+A_t^\frac{1}{3}}{Z_p Z_t}$ is a positive dimensionless parameter. Note that the parameter $x$, which appears in the argument of the logarithm of the above equation, can also be written as $x = \exp({\frac{R_B-r}{a}})$ and is larger than one in the most cases. The barrier potential $V_B$ is given by
\begin{equation}
    V_B = \frac{Z_p Z_t e^2}{4\pi\epsilon_0 R_B} - \frac{15}{x+1}
    \label{eq49}
\end{equation}
\subsection{Moustabchir-Royer (MR) formula}
{ Moustabchir and Royer \cite{moustabchir2001analytic} proposed two formulas for  the fusion barrier height and radius from
a fitting procedure on generalised liquid drop model \cite{royer1984fission} data on a large number of fusion reactions as follows:
\begin{align}
    V_B=-19.38+\frac{2.1388Z_pZ_t+59.427(A_p^{1/3}+A_t^{1/3})-27.07\ln\left(\frac{Z_pZ_t}{A_p^{1/3}+A_t^{1/3}}\right)}{(A_p^{1/3}+A_t^{1/3})(2.97-0.12\ln(Z_pZ_t))}\label{rmvb}
    \end{align}
\begin{align}
    R_B=(A_p^{1/3}+A_t^{1/3})\left[1.908-0.0857\ln(Z_pZ_t)+\frac{3.94}{Z_pZ_t}\right]
\end{align}    
 The rms deviations were found to be 0.15 MeV and 0.08 fm in $V_B$ and $R_B$, respectively.} 
\section{Results and discussions}
\begin{figure*}
\centering
\includegraphics[height=22cm,width=\linewidth,]{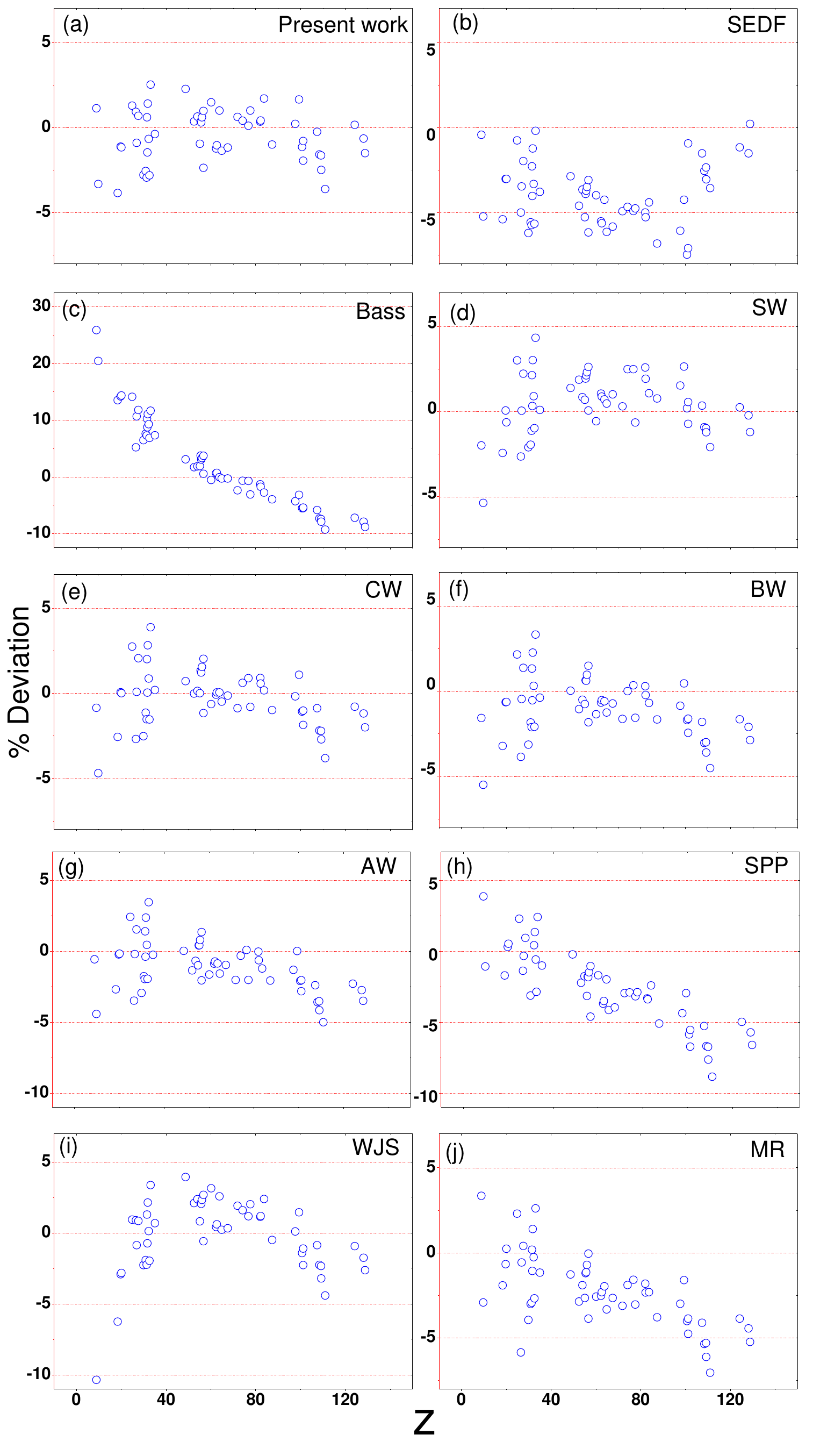} 
\caption{Comparison of percentage of deviation of different theoretical models from the experimental fusion barriers as a function of $z$.}
\label{f2}
\end{figure*}

\begin{figure*}
    \centering
    \includegraphics[width=\textwidth,height=20cm]{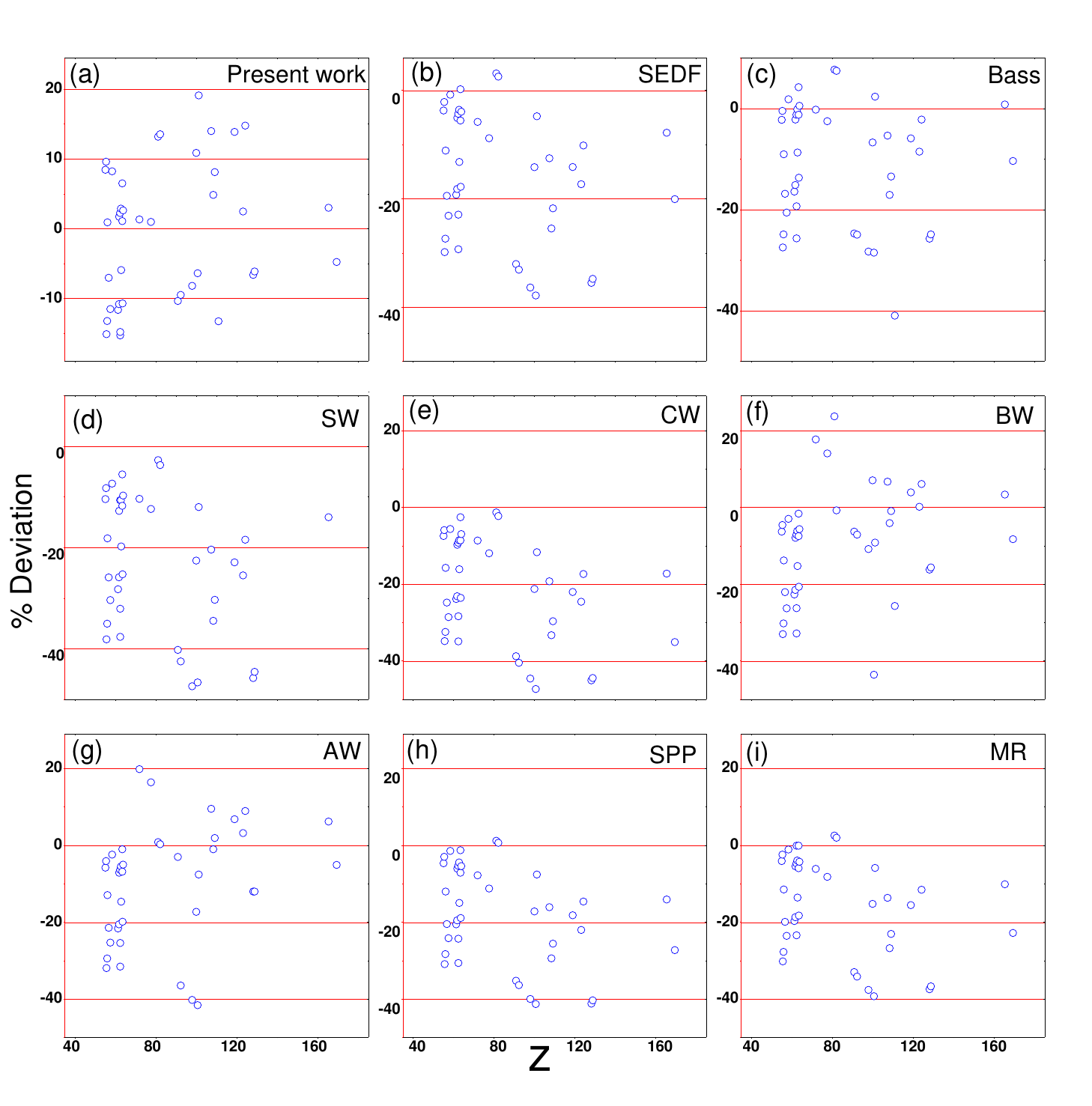}
    \caption{Comparison of percentage of deviation of different models from the experimental barrier radius as a function of $z$}
    \label{f3}
\end{figure*}

\begin{figure*}
\centering
\includegraphics[width=\linewidth,]{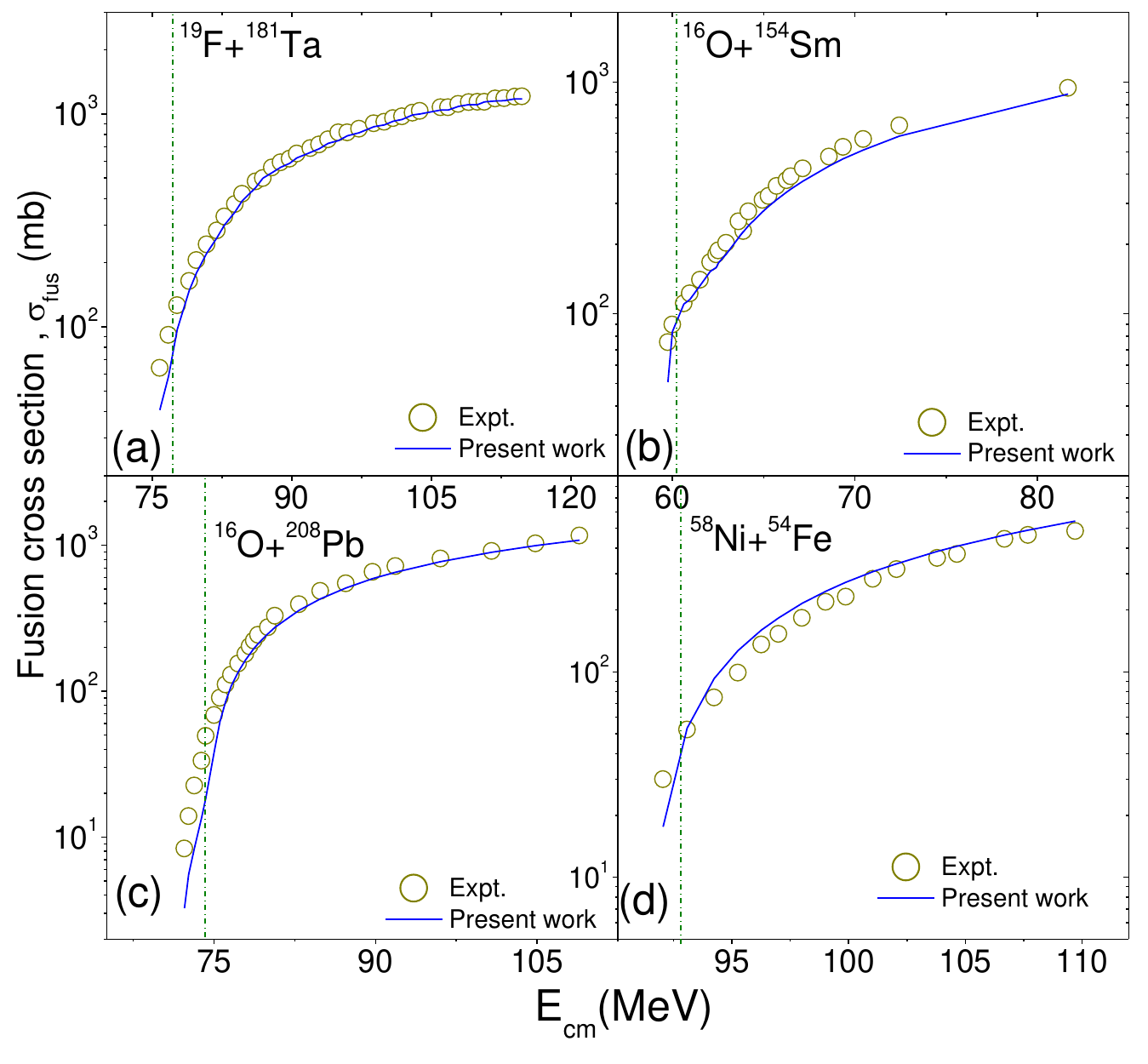} 
\caption{Comparison of total fusion cross section between theory and experiment as a function of $E_{cm}$ for the reactions (a) $^{19}$F+$^{181}$Ta \cite{shaikh2018investigation} and (b) $^{16}$O+$^{154}$Sm \cite{newton2004systematic}(c) $^{16}O+^{208}Pb$\cite{morton1999coupled} (d)$^{58}Ni+^{54}Fe$. The dashed vertical line indicates the fusion barrier for the corresponding reaction.}
\label{f4}
\end{figure*}

We have constructed the empirical formulae for the fusion and interaction barriers using the experimental results available in the literature as mentioned above. The fusion barriers $B_{fu}$ can be obtained from the equations \ref{eq4} and \ref{eq5},  and fusion barrier radius $R_B$  is from equation \ref{eq7}. The $B_{fu}$ values obtained from various theoretical models for different systems are compared with the experimental results in the rage of $9\leq z \leq 129$, which are not used to construct the present fusion barrier formula,  in Table \ref{t3}. Similar comparisons have also been done for $R_B$ in Table \ref{t4} for the range $55\leq z \leq 170$. To check which model gives the best agreement with the experimental results, the differences of both $B_{fu}$ and  $R_B$ values between the experiment and a specific model have been plotted as a function of $z$ in Fig. \ref{f2} and \ref{f3}, respectively. From the figure \ref{f2}, it is seen that the present empirical formula (Fig\ref{f2}(a)) is within $\pm 4\%$, most of the SEDF data overestimates up to $7\%$ (Fig.\ref{f2}(b)), Bass model underestimates for $z \leq 60$ and overestimates for $z \geq 60$ (Fig.\ref{f2}(c)), SW predicts within $\pm 5\%$ (Fig.\ref{f2}(d)), similar comparison is found with CW data (Fig.\ref{f2}(e)), BW data (Fig.\ref{f2}(f)) and AW data (Fig.\ref{f2}(g)), while SPP data shows about an underestimation for $z$ up to 40 and overestimation up to 10\% for $z>$40. On the other hand, Fig.\ref{f3} gives a clear impression that the deviation between the experiment and theory for the fusion barrier radius is much larger than the fusion barrier. Whatsoever, we can find that the present empirical formula (Fig\ref{f3}(a)) is within $\pm$20\%, deviations of the SEDF data are within $\substack{+40\% \\ -5\%}$ (Fig\ref{f3}(b)), Bass model estimates within  $\substack{+40\% \\ -6\%}$
 (Fig.\ref{f3}(c)), SW predicts within $\substack{+50\% \\ -0\%}$ (Fig.\ref{f3}(d)), similar deviation is found with CW (Fig.\ref{f3}(e)), BW data (Fig.\ref{f3}(f)) and AW data show about $\substack{+40\% \\ -20\%}$ (Fig.\ref{f3}(g)) deviations, while SPP data shows an overestimation of 40\%. Further, to have a quantitative evaluation on the model predictions we have obtained the sum of the squared residuals ($SSR= \displaystyle\sum_{i=1}^{n} e_i^{2}$), where $e_i$ is the $i^{th}$ residual or difference and n is the number of data points. The mean squared errors ($\sigma_\epsilon^2 = \frac{SSR}{n-2}$) are shown in Table \ref{t5}. The lowest mean squared error on both the fusion barrier and radius is obtained for the the present model, thus the best available formula till date. Note that the empirical formula of SJW \cite{swiatecki2005fusion} does not predict fusion barrier radius, it is considered to be disqualified in this test. The BW model is the second best and the AW model is the third best in the region considered here.\\
 \indent  We can see from the Fig. \ref{f1} that very few data are existing for $z > 150$, because the reaction mechanism at this region is complex and the data analysis is a tricky affair. If the experimental apparatus is powerful enough in the sense of measuring a number of physical parameters, right kind of data analysis can then be possible. Often such apparatus is not available, thereby appropriate analysis method cannot be straightforwardly chosen as discussed below. Therefore, available experimental results are very limited. Whatsoever, the available experimental results for the heavy ion reactions aiming to form actinide and transactinide elements have been compared with the present formula, WJS formula \cite{swiatecki2005fusion}, W.D. Myers (WDM) model \cite{myers2000nucleus} and H.C. Manjumatha (HCM) formula \cite{manjunatha2018pocket} in the Table \ref{t6}. The  mean square errors in the present formula, WDM model, WJS formula, and HCM formula are found to be 2.26, 4.98, 4.42 and 5.72, respectively. Hence, the predictions from the present formula is the best of all. An important point to note that excellent agreement is found even with very precision experiments just conducted recently \cite{banerjee2019mechanisms}. Hence, the present formula is an appropriate formula that may be very useful for synthesizing the superheavy elements [discussed later]. This formula is thus useful for the entire region of $8\leq z \leq 286$. Of course, for the region of $0 < z \leq 8$ where often nuclear astrophysics experiments \cite{kappeler1998current} are conducted, the BW model can be used.\\
{ \indent Very recently, Ganiev and Nasirov \cite{ganiev2020comparative} discussed predictive power of various nucleus-nucleus potentials. The proximity potential \cite{blocki1977proximity} and its use in AW model \cite{winther1995dissipation} give deviation more than 4 MeV from the experimental values of $V_B$ for several fusion reactions. More importantly, the difference increases for the reactions with massive nuclei including the mostly used Bass potential \cite{bass1973threshold, bass1974fusion,bass1977nucleus,bass1980theory} of the proximity potential. For some fusion systems, the results of CW \cite{christensen1976evidence} and BW model \cite{broglia1991heavy} are close to the experimental data, but these models cannot describe the heavy-ion reactions such as ${86}Kr+^{208}Pb$. In contrast, predictions of a double-folding model (DFM) \cite{ganiev2020comparative} based on density-dependent Migdal potential \cite{migdal1959superfluidity} are closed to all the experimental values including the heavy-ion reactions. We compare the present results for the $V_B$ with this DFM model along with the DFM using M3YReid interaction with zero-range exchange part \cite{bertsch1977interactions} and the Paris CDM3Y3 interaction with finite-range exchange part \cite{anantaraman1983effective} and experimental data obtained from the cited references in Table \ref{comparison-DFM}. The mean squared error analysis give a clear impression that predictions from our formula are the best of all these models.} \\
\indent Synthesis of superheavy elements by heavy-ion reactions is a complex problem because of quasi-fission etc. A fully equilibrated compound system needs to be formed to produce the new elements. This depends on entrance channel parameters \cite{manjunath2020entrence}, deformation parameters, orientation, internal excitation, transfer and initial kinetic energy of the projectile \cite{manjunath2020Quasifission}, which can be picked by the knowledge of theoretical $B_{fu}$. {However, the barrier distributions of many heavy-ion reactions used for synthesis of superheavy nuclei shows double hump behavior, for example, less mass asymmetry cold fusion reactions. Out of the two humps, the inner peak is the higher. While the warm fusion reactions, due to higher mass asymmetry, exhibit no double hump barriers \cite{royer2002formation}. At a given mass
asymmetry up to about $\eta$ = 0.5 the potential barrier exhibits a two hump shape, but for larger $\eta$ it displays only one hump \cite{poenaru2016spontaneous}. Hence, double hump problem appear only for the cold fusion type of reactions only and the present predictions might represent the higher potential barrier of the corresponding distribution for such reactions. Such scenario has been discussed later by several authors time to time, for example, a very recent article by Royer $et. al.$ \cite{royer2020fusion}. Hence, a precisely known value from this present formula will thus promote us to select the kinetic energy judiciously. Recently, an reaction contact time experiment \cite{albers2020zeptosecond} has made use of the WJS formula \cite{swiatecki2005fusion} (not the WDM model \cite{myers2000nucleus}, although it is quoted there). We showed just above that uncertainty in the WJS formula \cite{swiatecki2005fusion} is much higher than the present formula. The fusion barriers of the reactions aiming for the superheavy nucleus $=120$ have also been compared with different formulae in Table \ref{t6}}.\\
\indent Though the present empirical formula is not specific to a particular form of the nucleus-nucleus potential, but the second as well as the third best models are based on the Woods-Saxon potential (equation (\ref{eq23}) and characterised by the three potential parameters including the potential height parameter $V_0$, the diffuseness parameter $a_0$ and the radius parameter $r_0$. The present formula predicts the fusion barrier and radius pretty well, if we substitute these values in the three equations \ref{eq3} and solve  for these three parameters for a reaction, we can constitute the concerned nucleus-nucleus potential quite well. Such an exercise is followed for several reactions and the Woods-Saxon potential parameters are shown in Table \ref{t7} and compared with the potential parameters used in the earlier analysis, where mostly the Akyuz Winther parameters \cite{akyuzparameterization} have been used \cite{scobel1976fusion}. Sometimes the parameters are altered to have an agreement with the fusion cross section measurements \cite{shaikh2018investigation}. Even at times, modified Akyuz Winther parameters are chosen by fixing the potential height at 100 MeV \cite{newton2004systematic}. Note that in some models another parameter is  kept fixed, for instance, the $a_0$ value is fixed for the BW and AW models to 0.63 fm. In contrast, we have kept all the parameters free while the basic equations are solved (equation \ref{eq3}). The $V_0, a_0$ and $r_0$  span over 26-135 MeV, 0.92-1.76 fm, and 0.85-1.24 fm, respectively. In contrast, earlier Akyuz Winther parameters for $V_0, a_0$ and $r_0$  fall in the range of 76-105 MeV, 0.67-1.06 fm and 1.06-1.19 fm, respectively. \\
\indent Once we solve for the potential parameters $V_0, a_0$ and $r_0$ by above prescriptions and assume that the tail of the nuclear potential to be exponential with diffuseness $a_0$, one can readily
obtain $R_B$ from \cite{rowley1989scaling}
\begin{align}
V_0=\frac{Z_1Z_2e^2}{R_B}(1-\frac{a_0}{R_B}). 
\end{align} 
We can obtain now another important parameter called the barrier width $(\hbar\omega)$ from the relation given by Rowley $et~al.$ \cite{rowley1991distribution} as follows
\begin{align}
(\hbar\omega)^2=\frac{\hbar^2V^{\prime\prime}(R_B)}{\mu}= \frac{Z_1Z_2e^2\hbar^2}{\mu R_B^2}(\frac{1}{a_0}-\frac{2}{R_B}). 
\end{align} 
Use of these barrier radius and width in the Wong's phenomenological formula \cite{wong1973interaction} gives us the fusion cross section as follows
\begin{align}
\sigma_{fu}^{Wong}=\frac{\hbar\omega R_b^2}{2E}ln[1+exp\frac{2\pi(E-V_B)}{\hbar\omega}]. 
\end{align} \\
\indent We made an attempt to check whether the predicted fusion barrier and radius from the present formula can reproduce the measured excitation fusion i.e., fusion cross section as a function of centre of mass energy of the projectile. It is done by a direct substitution of $R_B$ and $\hbar\omega$ in the classic Wong formula \cite{wong1973interaction} given above for many reactions. As for example, the Fig.\ref{f4} shows a comparison for four reactions $^{19}$F+$^{181}$Ta \cite{shaikh2018investigation}, $^{16}$O+$^{208}$Pb, $^{16}$O+$^{154}$Sm, and $^{58}$Ni+$^{54}$Fe \cite{newton2004systematic}. The comparison of the total fusion cross section between the phenomenological formula \cite{wong1973interaction} and experiment displays a very good agreement for $E_{cm} > B_{fu}$ and a departure starts appearing at low energies $E_{cm} < B_{fu}$, except for the reaction $^{16}$O+$^{154}$Sm, because of the strong channel coupling effects for some reactions in the sub-barrier region \cite{hagino2012subbarrier}. The present scenario reiterates the fact that no single, energy-independent potential can simulate the fusion cross sections at the sub-barrier region, where the well-known influence of the coupling effects is vital. {A recent study \cite{gharaei2019assessment} has discussed the sub-barrier fusion properties on a  different view point, which shows importance of the projectile mass and surface energy coefficients in heavy-ion fusion at sub-barrier energies. Further, it finds essence of fusion Q-value rule \cite{stefanini2007sub} for some fusion systems.} \\
\indent One point may be worth mentioning that owing to deformation in the projectile and/or target nuclei, a distribution of barrier heights is observed in the experiments. The fusion barrier height from the present empirical formula represents an average value in the case of such distribution. Obviously, the barrier width shows also an average value and its value for certain reactions are compared with other models in Table \ref{t7}.\\
\indent In recent years, microscopic mechanisms which can impact the nucleus-nucleus potential has been studied in the frame work of time dependent Hartree-Fock (TDHF) approach \cite{negele1982mean} to provide a rather unique tool for describing nuclear structure and nuclear reactions over the whole nuclear
chart. Assuming the densities of the target and projectile remain constant and equal to their respective ground state densities, this leads to the so-called frozen density TDHF (FD-TDHF) approximation. Washiyama and Lacroix \cite{washiyama2008energy} consider a different approach based on a macroscopic reduction of the mean-field dynamics, called dissipative-dynamics in TDHF (DD-TDHF). We have compared the DD-TDHF predictions for the fusion barrier and barrier radius for several reactions with our predictions and available experimental values in Table \ref{t8}, and noticed a good agreement. In recent years, Simenel and his co-workers  \cite{simenel2010particle,simenel2013microscopic} and Yilmaz $et.al.$ \cite{yilmaz2011nucleon} have applied TDHF without any approximation to verify the FD-TDHF and DD-TDHF predictions in several cases and found good accordance. {Mean squared error is given in the bottom row of the table, which implies that the DD-TDHF predictions with low E$_{CM}$ are the best if we judge on both $V_B$ and $R_B$ simultaneously.}\\
\indent Let us discuss briefly the data analysis issue of the heavy ion reactions here.  When a projectile collides with a target nucleus near the fusion barrier energy, besides the evaporation residues (ER) and binary fission \cite{klotz1989properties}, the quasi-fission \cite{Swiatecki1980Three,swiatecki1981dynamics} is another process that contributes considerably in the reaction cross sections. Furthermore, quasi-fission occurs before the target and projectile fuse into a compound nucleus not only for the heavy systems $z_t.z_p > 1600$ but also in much lighter systems $z_t.z_p\approx 800$ \cite{rafiei2008strong},  which results in the hindrance of the formation of ER from the equilibrated compound nucleus.  The fusion-fission can also take place from an incomplete fusion reaction \cite{diaz2002effect,diaz2007relating}, in which only a part of the projectile fuses with the target and  the incompletely fused binary system equilibrates in the compound nucleus. Often the quasi-fission products are considered as the products of the deep-inelastic collisions \cite{broglia1974deep} in the experimental data analysis and their contribution is obviously not included in the capture cross section, the estimation of the fusion probability from such analysis appears unreliable. If the effect of quasi-fission and incomplete fusion-fission reaction are not considered in the analysis the reaction cross sections will certainly be erroneous.  Such circumstances may of course distort the value of the fusion barrier if the contribution of the quasi-fission and incomplete fusion-fission reaction vary with the beam energy. However, a recent thorough study shows that the cross section measured from the fission products $\sigma_{fis}$ could be well reproduced by scaling the capture cross section $\sigma_{cap}$ for a cold fusion by constant factors of 0.75 for $^{48}$Ca, 0.48 for $^{50}$Ti, and 0.22 for $^{54}$Cr projectiles on $^{208}$Pb target \cite{banerjee2019mechanisms}. Hence, the present empirical formula is suitable to realistically predict the fusion barrier for any new reactions planned in search for the formation of the superheavy elements.\\
\indent Let us now discuss the interaction barriers, which can be obtained from equation (\ref{eq6}) and compared with the experiments and model predictions. The experimental values available have been compared with the present empirical formula in Table \ref{t9}. These experimental values compare far better with the present formula than the Bass model. Mean squared error turns out to be 3.0 for the present formula, whereas that is  16.0 for the Bass model.  Hence, the present model of the interaction barriers is also recommended for future applications for $59\leq z\leq 313$ and the Bass model for $0 < z\leq 59$. \\ 
\indent  As discussed in the introduction, the motivation of the present work was based on a work \cite{sharma2017shakeoff} that the observed resonance like structures with the projectile x-ray energies appeared at a certain energy closure to the fusion barrier energy even though the projectile x-ray production mechanism is close to the elastic collision as examined by the Bass model \cite{bass1974fusion,bass1973threshold}. The scenario remains unchanged with the refined fusion and interaction barrier energies for the systems $^{12}$C($^{56}$Fe,$^{56}$Fe), $^{12}$C($^{58}$Ni,$^{58}$Ni) and $^{12}$C($^{63}$Cu,$^{63}$Cu). The physical reason for this anomaly is identified as a process called the nuclear orbiting resonance (dinuclear complexes) \cite{Braun}. Detail of this aspect is out of the scope of this work and will be published elsewhere. \\
\section{Conclusion}
In this paper, using the experimental values available in the literature, empirical formulae for the fusion and interaction barriers have been obtained. The experimental values available  for the fusion barrier radius give us an option to find a formula for the fusion barrier radius also.  The present study is restricted to the fusion and interaction barriers for the reactions in the regime $8\leq z \leq286$ and $59\leq z \leq313$, respectively.  We have carried out a comparative study of the fusion barriers as well as barrier radius between the present empirical formula and various empirical, semi-empirical models, and microscopic theories along with the experimental results. According to a thorough comparison with the experimental values, the present formula is  found to be the best of all the models considered in this study for the said regions. Further, to examine its predictability, the fusion barrier and barrier radius have been used in the classic Wong fusion cross section formula and the total fusion cross sections are found to be compared well with the experimental values. The fusion barrier and barrier radius obtained from the present formulae are used to solve for the potential parameters of the Woods-Saxon potential using the three basic equations \ref{eq3}. The parameters so obtained are quite different from the ones used in the earlier studies. \\
\indent The present fusion barrier formula is showed an excellent accordance with the experimental data in Table \ref {t6}. This comparison includes mostly the heavy ion reactions aiming to synthesize the superheavy elements. Further, the agreement between the experimental and present interaction barrier formula is pretty good. Hence the present fusion  and interaction barrier formulae together can be confidently used for planning experiments for the synthesis of the new superheavy elements. Similarly, current interaction barriers have been compared with the experimental interaction barriers quite well. Though we revealed in this work that the fusion and interaction barrier from  the present work can be estimated very well, but it is an empirical formula. { A refined and complete theoretical model such as  DD-TDHF \cite{washiyama2008energy} is found to be even better than the present formula. More such theoretical calculations are highly desirable to shed better light on the reaction mechanism of  the heavy-ion collisions. At the same time, a benchmark experiment is of high demand that takes care of quasi-elastic events, deep inelastic collisions, capture reactions, etc. for the heavy-ion reactions, in a proper manner.} \\ 
\textbf{Acknowledgements:}
We would like to acknowledge the illuminating discussions with Subir Nath, Ambar Chatterjee, B.R. Behra and S. Kailas. \\

For corresponding: $\#$ nanditapan@gmail.com and
$*$ manjunathhc@rediffmail.com \\
\bibliographystyle{apsrev4-2}
\bibliography{apssamp.bib}
%
\newpage
\begin{center}
\begin{longtable}{p{4cm}p{3cm}p{1.3cm}}
\caption{The fusion barriers $(B_{fu})$ for the following two body systems have been used in Fig 1(a) to obtain the empirical formula for estimating the $B_{fu}$ for any system in the bound $8\leq z \leq286$.}
\\\hline
\hline
System & $z$ & $B_{fu} (MeV)$   \\ \hline \\
$^{12}$C+$^{15}$N & 8.83 &  6.80\cite{vaz1981fusion}\\              
$^{12}$C+$^{16}$O & 9.98 & 7.50\cite{sperr1976oscillations}\\ 
$^{12}$C+$^{26}$Mg & 13.71& 11.5\cite{jachcinski1981fusion}\\ 
$^{12}$C+$^{30}$Si & 15.57 & 13.2\cite{vaz1981fusion}\\
$^{16}$O+$^{27}$Al & 18.84 & 43.6\cite{Eisen1977Total}\\ 
$^{24}$Mg+$^{26}$Mg & 24.63 & 20.8\cite{gary1982fusion}\\
$^{26}$Mg+$^{32}$S & 31.28 & 27.5\cite{berkowitz1983fusion}\\
$^{12}$C+$^{92}$Zr & 35.27 & 32.3\cite{newton2001experimental}\\
$^{16}$O+$^{72}$Ge & 38.32 & 35.4\cite{aguilera1986search}\\
$^{32}$S+$^{40}$Ca & 48.52 & 43.3\cite{zanganeh2015analytical}\\
$^{48}$Ca+$^{48}$Ca & 55.03 & 51.7\cite{zanganeh2015analytical}\\
$^{27}$Al+$^{70}$Ge & 58.42 & 55.1\cite{zanganeh2015analytical}\\
$^{32}$S+$^{58}$Ni & 63.58 & 59.5\cite{gutbrod1973fusion}\\ 
$^{40}$Ar+$^{58}$Ni & 69.13 & 66.32\cite{zanganeh2015analytical}\\
$^{37}$Cl+$^{73}$Ge & 72.42 & 69.20\cite{zanganeh2015analytical}\\
$^{40}$Ca+$^{62}$Ni & 75.90 & 72.3\cite{vaz1981fusion} \\
$^{32}$S+$^{89}$Y & 81.68 & 77.8\cite{mukherjee2002dominance}\\
$^{16}$O+$^{238}$U & 84.43 & 80.8\cite{zanganeh2015analytical}\\
$^{28}$Si+$^{120}$Sn & 87.84 & 85.9\cite{zanganeh2015analytical}\\
$^{48}$Ca+$^{96}$Zr & 97.41 & 95.9\cite{stefanini2006fusion}\\
$^{40}$Ca+$^{96}$Zr & 100.01 & 93.6\cite{gutbrod1973fusion}\\ 
$^{40}$Ar+$^{121}$Sb & 109.72 & 111\cite{gauvin1974nuclear}\\ 
$^{40}$Ca+$^{124}$Sn & 118.95 & 113\cite{scarlassara2000fusion}\\ 
$^{28}$Si+$^{198}$Pt & 123.18 & 121\cite{nishio2000fusion}\\
$^{40}$Ar+$^{154}$Sm & 127.11 & 121\cite{reisdorf1985fusion}\\ 
$^{40}$Ar+$^{165}$Ho & 135.43 & 141.4\cite{zanganeh2015analytical}\\
$^{40}$Ca+$^{192}$Os & 165.42 & 168.1\cite{zanganeh2015analytical}\\
$^{84}$Kr+$^{116}$Cd &186.67 & 204\cite{gauvin1972complete}\\
$^{74}$Ge+$^{232}$Th & 278.45 & 310\cite{gauvin1972complete}\\
$^{86}Kr+^{208}Pb$ & 285.52 & 299.00\cite{dutt2010systematic}\\  
\hline
\hline\label{t1}
\end{longtable}
\end{center}
\newpage
\begin{table}
\centering
\caption{The interaction barriers $(B_{int})$ for the following two body systems have been used in Fig \ref{f1} (b) to obtain the empirical formula for estimating the $B_{int}$ for any unknown systems in the range $59\leq z \leq313$.}
\begin{tabular}{p{4cm}p{3cm}p{1.3cm}}
\hline\hline
System & $z$ & $B_{int (MeV)}$   \\ \hline  \\   
$^{12}$C+$^{205}$Tl & 59.37 & 56.0\cite{le1972nuclear}\\ 
$^{12}$C+$^{209}$Bi & 60.55 & 57.0\cite{le1972nuclear}\\ 
$^{12}$C+$^{238}$U & 65.04 & 62.2\cite{viola1962total}\\ 
$^{14}$N+$^{238}$U & 74.82 & 73.4\cite{viola1962total}\\ 
$^{16}$O+$^{205}$Tl & 76.99 & 77.0\cite{le1972nuclear}\\ 
$^{16}$O+$^{238}$U & 84.43 & 82.5\cite{viola1962total}\\ 
$^{20}$Ne+$^{238}$U & 103.23 & 102\cite{viola1962total}\\ 
$^{40}$Ar+$^{164}$Dy & 133.57 & 135\cite{le1971behavior}\\ 
$^{40}$Ar+$^{238}$U  & 172.19 & 171\cite{bass1973threshold}\\ 
$^{48}$Ti+$^{208}$Pb & 188.72 & 190.1\cite{mitsuoka2007barrier}\\
$^{54}$Cr+$^{208}$Pb & 202.78 & 205.8\cite{mitsuoka2007barrier}\\
$^{56}$Fe+$^{208}$Pb & 218.65 &223\cite{mitsuoka2007barrier}\\
$^{58}$Ni+$^{208}$Pb & 234.38 & 236\cite{mitsuoka2007barrier}\\
$^{70}$Zn+$^{208}$Pb & 244.86 & 250.6\cite{mitsuoka2007barrier}\\
$^{84}$Kr+$^{232}$Th & 307.86 & 332\cite{bass1973threshold}\\ 
$^{84}$Kr+$^{238}$U  & 313.14 & 333\cite{moretto1972shell}\\ 
\hline
\hline
\end{tabular}
\label{t2}
\end{table}
\newpage
\begin{center}\tiny
\begin{longtable*}{p{2cm} p{1.3cm} p{1.8cm} p{1.2cm} p{1.1cm} p{1.1cm} p{1.1cm} p{1.1cm} p{1.1cm} p{1.1cm} p{1.1cm}p{1.1cm}p{1.2cm}c}
\caption{Comparison of the fusion barrier $B_{fu}$ for different systems with experimental as well as various theoretical models. The reactions are listed in order of the increasing value of the $z$ parameter. The values for our model is taken from the equation (\ref{eq5}). }    \\\hline \hline
    ${System}$&$~~~z$&$Expt.[Ref.]$&$SEDF$&$Bass$&$SW$&$CW$&$BW$&$AW$&$SPP$&WJS& MR&$Pres.$&   \\
    \hline
$^{12}$C+$^{14}$N &8.93 &7.00\hspace{2mm} \cite{vaz1981fusion}&$7.03$&$5.19$& $7.14$&$7.06$&$7.11$&$7.04$&$6.73$&7.72& 6.77 &$7.47$     \\
$^{12}$C+$^{18}$O & 9.77 & 7.45\hspace{2mm} \cite{sperr1976fusion}&$7.84$&$5.93$& $7.85$&$7.80$&$7.86$&$7.78$& $7.53 $&8.46& 7.67 &$7.69$\\
 $^{12}$C+$^{17}$O& $9.88$ & $8.20$ \hspace{0.5mm} \cite{zanganeh2015analytical}&$7.97$&$5.97$& $7.86$&$7.89$&$7.94$&$7.86$& $7.59$&8.54& 7.68 &$7.78$   \\ 
 $^{20}$Ne+$^{20}$Ne& $18.42$ & $15.20$ \cite{shapira1983fusion}&$16.02$&$13.15$& $15.57$&$15.59$&$15.69$&$15.61$& $15.46$&16.15 & 15.49 &$15.78$   \\
$^{18}$O+$^{28}$Si& $19.79$ &$16.90$\hspace{1mm}\cite{zanganeh2015analytical} &$17.41$&$14.51$&$16.89$&$16.89$&$17.01$&$16.94$&$16.85$&17.39& 17.01 &$17.09$   \\
 $^{16}$O+$^{28}$Si&$20.16$ &$17.23$ \cite{zanganeh2015analytical}&$17.75$&$14.76$&$17.34$&$17.23$&$17.34$&$17.26$&$17.14$&17.71& 17.19 &$17.43$\\
 $^{24}$Mg+$^{24}Mg~$&$24.96$ &$22.30$ \cite{jachcinski1981fusion}&$22.47$&$19.16$& $21.63$&$21.69$&$21.82$&$21.76$&$21.79$&22.09& 21.79 &$22.01$   \\
 $^{6}$Li+$^{144}$Sm&$26.35$ &$24.65$ \cite{zanganeh2015analytical}&$24.65$&$22.22$& $24.19$&$24.32$&$24.47$&$24.33$&$23.78$&23.36&24.68 &$23.35$   \\
 $^{7}$Li+$^{159}$Tb&$26.60$ &$23.81$ \cite{zanganeh2015analytical}&$25.00$&$22.58$&$24.44$&$24.45$&$24.73$&$24.64$&$24.14$ &23.59&25.21 & $23.59$   \\
  $^{16}$O+$^{40}$Ca&$26.94$ &$23.70$ \cite{jachcinski1981fusion}&$24.52$&$21.18$&$23.69$&$23.68$&$23.81$&$23.75$&$23.78$& 23.91&23.84 &$23.91$   \\
 $^{26}$Mg+$^{30}$Si&$27.68$ &$24.80$ \cite{zanganeh2015analytical}&$25.29$&$21.88$&$24.25$&$24.29$&$24.46$&$24.42$&$24.57$& 24.59&24.70 &$24.62$   \\
 $^{14}$N+$^{59}$Co&$29.99$ &$26.13$ \cite{gomes1991fusion}&$27.75$&$24.45$&$26.68$&$26.79$&$26.95$&$26.90$&$26.95$&26.72&27.16 &$26.86$   \\
 $^{26}$Mg+$^{34}$S&$30.96$ &$27.11$ \cite{berkowitz1983fusion}&$28.62$&$25.06$&$27.64$&$27.42$&$27.61$&$27.59$&$24.07$&27.63&27.93 &$27.81$   \\
$^{24}$Mg+$^{34}$S&$31.35$ &$27.38$ \cite{berkowitz1983fusion}&$28.95$&$25.39$&$27.69$&$27.80$&$27.96$&$27.92$&$24.37$&27.99&28.18 &$28.19$   \\
 $^{30}$Si+$^{30}$Si&$31.53$ &$28.54$ \cite{aguilera1986search}&$29.19$&$25.62$&$27.93$&$27.97$&$28.16$&$28.14$&$28.42$&28.17&28.49 &$28.37$   \\
 $^{24}$Mg+$^{32}$S&$31.69$ &$28.10$ \cite{berkowitz1983fusion}&$29.23$&$25.65$& $28.01$&$28.09$&$28.25$&$28.21$&$24.61$&28.31&28.40 &$28.51$   \\
 $^{6}$Li+$^{208}$Pb&$31.77$ &$30.10$\hspace{1mm}\cite{zanganeh2015analytical} &$30.27$&$28.20$&$30.12$&$29.97$&$30.18$&$30.06$&$29.40$&28.39& 30.54 &$28.59$   \\
 $^{28}$Si+$^{30}$Si&$31.90$ &$29.13$ \cite{gary1982fusion}&$29.49$&$25.91$&$28.25$&$28.31$&$28.47$&$28.44$&$28.74$&28.51&28.72 &$28.72$   \\
$^{28}$Si+$^{28}$Si&$32.27$ &$28.89$ \cite{gary1982fusion}&$29.85$&$26.22$&$28.63$&$28.64$&$28.80$&$28.76$&$29.06$&28.85&28.97 &$29.08$   \\
 $^{20}$Ne+$^{40}$Ca&$32.60$ &$28.60$ \cite{zanganeh2015analytical}&$30.22$&$26.64$&$28.88$&$29.04$&$29.20$&$29.16$&29.16&$29.42$&29.37 &$29.41$   \\
 $^{24}$Mg+$^{35}$Cl&$33.14$ &$30.70$ \cite{zanganeh2015analytical}&$30.76$&$27.13$&$29.37$&$29.51$&$29.68$&$29.64$&29.66&$29.96$& 29.90 &$29.92$   \\
 $^{16}$O+$^{58}$Ni&$35.05$ &$31.67$ \cite{zanganeh2015analytical}&$32.87$&$29.36$&$31.64$&$31.61$&$31.79$&$31.75$&31.45&$31.99$& 32.04 &$31.79$   \\
 $^{12}$C+$^{152}$Sm&$48.78$ &$46.39$ \cite{zanganeh2015analytical}&$47.72$&$44.97$&$45.75$&$46.06$&$46.38$&$46.38$&44.56&$46.50$&46.98 &$45.34$   \\
$^{18}$O+$^{124}$Sn&$52.58$ &$49.30$ \cite{zanganeh2015analytical}&$51.57$&$48.48$&$48.38$&$49.31$&$49.82$&$49.97$&$50.40$&48.26&50.72 &$49.12$   \\
$^{16}$O+$^{116}$Sn&$54.08$ &$50.94$ \cite{zanganeh2015analytical}&$52.80$&$50.00$&$50.51$&$50.87$&$51.20$&$51.29$&$51.84$&49.72&51.91 &$50.61$   \\
$^{30}$Si+$^{64}$Ni&$55.15$ &$51.20$ \cite{stefanini1986heavy}&$53.90$&$50.25$&$50.85$&$51.20$&$51.59$&$51.71$&$52.82$&50.78&52.56&$51.69$   \\
 $^{30}$Si+$^{62}$Ni&$55.48$ &$52.20$ \cite{stefanini1986heavy}&$54.24$&$50.56$&$51.19$&$51.51$&$51.87$&$51.99$&$53.14$&51.11&52.82 &$52.02$   \\
 $^{28}$Si+$^{64}$Ni&$55.71$ &$52.40$ \cite{stefanini1986heavy}&$54.35$&$50.82$&$51.29$&$51.76$&$52.08$&$52.18$&$53.36$&51.32&53.00 &$52.24$   \\
$^{28}$Si+$^{62}$Ni&$56.04$ &$52.89$ \cite{stefanini1986heavy}&$54.74$&$51.14$&$51.67$&$52.07$&$52.38$&$52.47$&$53.68$&51.65&53.27 &$52.57$   \\
$^{40}$Ca+$^{48}$Ca&$56.70$ &$52.00$ \cite{zanganeh2015analytical}&$55.21$&$51.73$&$51.97$&$52.61$&$52.95$&$53.07$&$54.40$&52.31&54.01 &$53.23$   \\
$^{28}$Si+$^{58}$Ni&$56.75$ &$53.80$\hspace{1mm}\cite{stefanini1986heavy}&$55.46$&$51.82$&$52.39$&$52.71$&$53.00$&$53.08$&$54.37$&52.35&53.83 &$53.28$   \\
 $^{12}$C+$^{204}$Pb&$60.17$ &$57.55$ \cite{zanganeh2015analytical}&$59.84$&$57.89$&$57.88$&$57.92$&$58.33$&$58.50$&$58.53$&55.74&59.03 &$56.71$   \\
$^{40}$Ca+$^{48}$Ti&$62.37$ &$58.17$ \cite{sonzogni1998transfer}&$61.38$&$57.82$&$57.56$&$58.22$&$58.56$&$58.68$&$60.32$&57.92& 59.64 &$58.89$   \\
$^{35}$Cl+$^{54}$Fe&$62.69$ &$58.59$ \cite{zanganeh2015analytical}&$61.89$&$58.20$&$58.08$&$58.56$&$58.90$&$59.02$&$60.65$&58.24&59.95 &$59.21$   \\
$^{16}$O+$^{144}$Sm&$63.91$ &$61.03$ \cite{leigh1995barrier}&$63.62$&$61.08$&$60.60$&$61.00$&$61.39$&$61.56$&$62.25$&59.45&62.23 &$60.42$   \\
$^{37}$Cl+$^{64}$Ni&$64.92$ &$60.60$ \cite{zanganeh2015analytical} &$64.32$&$60.80$&$60.32$&$60.90$&$61.36$&$61.56$&$63.12$&60.47&62.62 &$61.43$   \\
 $^{46}$Ti+$^{46}$Ti&$67.54$ &$63.30$ \cite{zanganeh2015analytical}&$66.99$&$63.50$&$62.66$&$63.39$&$63.76$&$63.92$&$65.81$&63.11&64.98 &$64.05$   \\
 $^{16}$O+$^{186}$W&$71.95$ &$68.87$ \cite{leigh1995barrier}&$72.26$&$70.53$&$68.66$&$69.48$&$69.99$&$70.27$&$70.91$&67.55&71.02 &$68.44$   \\
$^{28}$Si+$^{92}$Zr&$74.16$ &$70.93$ \cite{newton2001experimental}&$74.24$&$71.43$&$69.17$&$70.50$&$70.93$&$71.16$&$72.99$&69.81&72.27 &$70.65$   \\
$^{40}$Ca+$^{58}$Ni&$76.81$ &$73.36$ \cite{vaz1981fusion}&$76.96$&$73.90$&$71.54$&$72.71$&$73.11$&$73.30$&$75.70$&72.51&74.52 &$73.29$   \\
$^{16}$O+$^{208}$Pb&$77.68$ &$74.90$ \cite{morton1999coupled}&$78.46$&$77.25$&$75.39$&$75.50$&$76.07$&$76.42$&$77.07$&7339&77.18 &$74.15$   \\
$^{48}$Ti+$^{58}$Ni&$82.08$ &$78.80$ \cite{vinodkumar1996absence}&$82.74$&$79.86$&$76.76$&$78.09$&$78.57$&$78.83$&$81.42$&77.91&80.24 &$78.53$   \\
$^{36}$S+$^{90}$Zr&$82.23$ &$79.00$ \cite{stefanini2000near}&$83.17$&$80.39$&$77.48$&$78.55$&$79.18$&$79.51$&$81.68$&78.11&80.85 &$78.68$\\
$^{19}$F+$^{197}$Au&$83.77$ &$81.61$ \cite{zanganeh2015analytical}&$85.20$&$83.88$&$80.73$&$81.47$&$82.18$&$82.61$&$83.58$&79.66&83.50 &$80.22$   \\
 $^{35}$Cl+$^{92}$Zr&$87.34$ &$82.94$ \cite{newton2001experimental}&$88.60$&$86.25$&$82.31$&$83.76$&$84.32$&$84.66$&83.35&$87.17$&86.09 &$83.77$   \\
 $^{35}$Cl+$^{106}$Pd&$97.70$ &$94.30$ \cite{zanganeh2015analytical}&$100.02$&$98.36$&$92.86$&$94.48$&$95.11$&$95.54$&$98.43$&94.21&97.13 &$94.11$   \\
$^{32}$S+$^{116}$Sn&$99.36$ &$97.36$ \cite{zanganeh2015analytical}&$101.50$&$100.47$&$94.79$&$96.31$&$96.92$&$97.35$&$100.24$&95.95&98.93 &$95.75$   \\
$^{58}$Ni+$^{60}$Ni&$100.69$ &$96.00$ \cite{zanganeh2015analytical}&$103.17$&$101.33$&$95.82$&$97.04$&$97.63$&$98.01$&$101.64$&97.36&99.93 &$97.09$   \\
$^{40}$Ca+$^{90}$Zr&$101.25$ &$96.88$ \cite{zanganeh2015analytical}&$103.75$&$102.26$&$96.34$&$97.87$&$98.45$&$98.86$&$102.26$&97.94& 100.63 &$97.65$   \\
$^{58}$Ni+$^{58}$Ni &101.27 &95.8\hspace{2mm}  \cite{timmers1998case}&$96.70$&$102.01$&$96.50$&$97.59$&$98.15$&$98.51$&$102.25$&97.97&100.36 &$97.67$ \\
$^{40}$Ar+$^{122}$Sn &107.40 &103.6  \cite{beckerman1980dynamic}&$105.18$&$109.70$&$103.24$&$104.51$&$105.47$&$106.09$&$109.07$&104.51&107.86 &$103.86$\\
$^{40}$Ar+$^{116}$Sn &108.47 &103.3  \hspace{0mm}\cite{reisdorf1985fusion}&$105.93$&$110.92$&$104.26$&$105.57$&$106.45$&$107.02$&$110.22$&105.63&108.84 &$104.94$\\
$^{40}$Ar+$^{112}$Sn &109.22 &104.0 \cite{reisdorf1985fusion}&$106.44$&$111.78$&$105.02$&$106.30$&$107.12$&$107.67$&$111.02$&106.44&109.52$105.71$\\
$^{64}$Ni+$^{74}$Ge &109.29 &103.2 \cite{reisdorf1985fusion}&$106.34$&$111.37$&$104.46$&$106.00$&$106.92$&$107.50$&$111.09$&106.51&109.52 &$105.77$\\
$^{58}$Ni+$^{74}$Ge & 111.04 & 106.8 \cite{beckerman1982sub}&$107.50$&$113.49$&$105.97$&$107.76$&$108.49$&$109.01$&$112.99$&108.39&$111.11$ &$107.56$\\ 
$^{34}$S+$^{168}$Er & 124.24 &121.5  \cite{hagino2004large}&$122.92$&$130.30$&$121.20$&$122.46$&$123.51$&$124.30$&$127.56$&122.63&126.11 &$121.31$\\
$^{40}$Ar+$^{148}$Sm &128.14 &124.7 \cite{reisdorf1985fusion}&$126.60$&$134.57$&$124.99$&$126.19$&$127.32$&$128.13$&$131.85$&126.88&130.25 &$125.45$\\
$^{40}$Ar+$^{144}$Sm &128.85 &124.4 \cite{reisdorf1985fusion}&$127.14$&$135.41$&$125.91$&$126.90$&$127.98$&$128.76$&$132.63$&127.66&130.92 &$126.28$\\
    \hline
\hline\label{t3}
\end{longtable*}
\end{center}
\newpage
\begin{center}
\begin{longtable} {p{1.9cm} p{1.4cm} p{1.6cm} p{1.3cm} p{1.3cm} p{1.3cm} p{1.3cm} p{1.3cm} p{1.3cm} p{1.3cm} p{1.3cm} p{1.1cm}c}   
\caption{Comparison of the fusion barrier radius $R_{B}$  for different systems with experimental as well as various theoretical models. The reactions are listed in order of the increasing value of the $z$ parameter. The values for our model is taken from the equation (\ref{eq7}). The superscript * indicates that the experimental radii for all the reactions are taken from the ones given in \cite{siwek2004empirical}.}
    \\\hline\hline
System&$z$&$Expt.^*$ & $SEDF$&$Bass$&$SW$&$CW$&$BW$&$AW$&$SPP$&MR &$Pres.$&   \\
    \hline \\
  $11.30$&$10.50$&$10.46$&$10.37$&$10.28$&$11.09$     \\
$^{30}$Si+$^{64}$Ni & 55.16 & 9.60&$9.96$&$9.82$& $10.60$&$10.32$&$10.21$&$10.16$& $10.05
$& 9.99 &$8.79$\\
 $^{30}$Si+$^{62}$Ni& $55.48$ & $9.70$&$9.91$&$9.75$& $10.50$&$10.28$&$10.15$&$10.10$& $9.99$&9.94 &$8.77$   \\ 
 $^{28}$Si+$^{64}$Ni& $55.71$ & $7.60$ &$9.87$&$9.69$& $10.50$&$10.25$&$10.11$&$10.03$& $9.95$& 9.90 &$8.75$   \\
  $^{28}$Si+$^{62}$Ni& $56.04$ & $7.70$ &$9.81$&$9.62$& $10.40$&$10.20$&$10.03$&$9.97$& $9.88$ & 9.84 &$8.72$   \\
  $^{30}$Si+$^{58}$Ni& $56.18$ &$8.80$ &$9.78$&$9.60$&$10.40$&$10.19$&$10.02$&$9.94$&$9.86$& 9.81 &$8.72$   \\
 $^{28}$Si+$^{58}$Ni&$56.75$ &$8.10$ &$9.68$&$9.47$& $10.20$&$10.11$&$9.89$&$9.84$&$9.76$&9.71 &$8.67$   \\
 $^{40}$Ca+$^{44}$Ca&$57.55$ &$7.90$ &$9.73$&$9.53$& $10.30$&$10.16$&$9.98$&$9.90$&$9.81$&9.76 &$8.81$   \\
 $^{40}$Ca+$^{40}$Ca&$58.48$ &$9.50$ &$9.58$&$9.33$&$10.20$&$10.04$&$9.79$&$9.73$&$9.64$ &9.61 &$8.72$   \\
  $^{36}$S+$^{64}$Ni&$61.35$ &$8.50$ &$10.14$&$9.90$&$10.90$&$10.53$&$10.43$&$10.34$&$10.25$&10.18 & $9.49$   \\
 $^{34}$S+$^{64}$Ni&$61.88$ &$8.50$ &$10.05$&$9.79$&$10.70$&$10.47$&$10.33$&$10.25$&$10.16$&10.09 & $9.42$   \\
 $^{40}$Ca+$^{50}$Ti&$61.94$ &$9.40$ &$9.88$&$9.61$&$10.60$&$10.32$&$10.15$&$10.07$&$9.97$&9.91 &$9.24$   \\
 $^{40}$Ca+$^{48}$Ti&$62.37$&$9.40$&$9.81$&$9.52$&$10.40$&$10.27$&$10.07$&$10.00$&$9.90 $&9.84 &$9.19$   \\
$^{32}$S+$^{64}$Ni&$62.44$ &$8.10$ &$9.96$&$9.67$&$10.70$&$10.40$&$10.23$&$10.16$&$10.07$&10.00 &$9.34$   \\
 $^{36}$Si+$^{58}$Ni&$62.46$ &$7.70$&$9.96$&$9.68$&$10.60$&$10.39$&$10.23$&$10.13$&$10.06$&10.09 &$8.84$   \\
 $^{40}$Ca+$^{46}$Ti&$62.83$ &$9.40$&$9.74$&$9.42$& $10.40$&$10.21$&$9.98$&$9.92$&$9.82$&9.77&$9.13$   \\
 $^{16}$O+$^{154}$Sm&$62.94$ &$9.60$ &$10.87$&$10.44$&$11.50$&$11.15$&$11.07$&$11.01$&$11.04$& 10.91 & $10.17$   \\
$^{17}$O+$^{144}$Sm&$63.49$ &$10.80$&$10.78$&$10.35$&$11.40$&$11.08$&$10.98$&$10.91$&$10.94$&10.81 &$10.10$   \\
 $^{16}$O+$^{148}$Sm&$63.51$ &$10.20$ &$10.77$&$10.33$&$11.40$&$11.08$&$10.97$&$10.90$&$10.93$&10.82&$10.09$   \\
 $^{32}$S+$^{58}$Ni&$63.59$ &$8.30$ &$9.78$&$9.44$&$10.40$&$10.26$&$10.02$&$9.95$&$9.87$&99.82 &$9.19$   \\
 $^{16}$O+$^{144}$Sm&$63.91$ &$10.30$ &$10.71$&$10.25$&$11.30$&$11.02$&$10.89$&$10.82$&$10.86$&10.47 &$10.03$   \\
 $^{16}$O+$^{186}$W&$71.95$ &$10.60$ &$11.22$&$10.63$&$11.70$&$11.52$&$8.73$&$8.50$&$11.43$&11.25 &$10.46$   \\
$^{16}$O+$^{208}$Pb&$77.68$ &$10.50$ &$11.43$&$10.77$&$11.80$&$11.76$&$9.03$&$8.78$&$11.68$&11.47 &$10.40$   \\
$^{36}$S+$^{96}$Zr&$81.21$ &$11.00$ &$10.66$&$10.16$&$11.30$&$11.15$&$8.40$&$10.91$&$10.87$& 10.72 &$9.55$   \\
$^{36}$S+$^{90}$Zr&$82.23$ &$10.80$ &$10.53$&$10.00$&$11.20$&$11.05$&$10.89$&$10.77$&$10.73$&10.59 &$9.34$   \\
 $^{36}$S+$^{110}$Pd&$90.94$ &$8.20$ &$10.83$&$10.23$&$11.50$&$11.38$&$8.72$&$8.45$&$11.09$&10.91 &$9.05$   \\
 $^{32}$S+$^{110}$Pd&$92.39$ &$8.00$ &$10.65$&$10.00$&$11.40$&$11.24$&$8.57$&$10.92$&$10.91$&10.7&$8.76$   \\
$^{64}$Ni+$^{64}$Ni&$98.00$ &$7.80$ &$10.64$&$10.01$&$11.50$&$11.28$&$8.65$&$10.94$&$10.92$&10.74 &$8.44$   \\
$^{40}$Ca+$^{96}$Zr&$100.01$ &$9.30$&$10.62$&$9.93$&$11.40$&$11.28$&$8.65$&$10.91$&$10.90$&10.72 &$8.29$   \\
$^{58}$Ni+$^{60}$Ni&$100.70$ &$7.50$&$10.34$&$9.64$&$11.00$&$11.05$&$10.77$&$10.62$&$10.60$&10.45 &$7.98$   \\
$^{40}$Ca+$^{90}$Zr&$101.25$ &$10.00$ &$10.48$&$9.77$&$11.20$&$11.17$&$10.92$&$10.76$&$10.76$&10.59 &$8.09$   \\
$^{40}$Ar+$^{122}$Sn&$107.40$ &$9.80$ &$11.03$&$10.33$&$11.80$&$11.69$&$9.15$&$8.87$&$11.38$&11.14 &$8.43$   \\
$^{40}$Ar+$^{116}$Sn&$108.47$ &$8.70$  &$10.92$&$10.19$&$11.70$&$11.60$&$9.06$&$8.79$&$11.26$&11.03 &$8.28$   \\
 $^{40}$Ar+$^{112}$Sn&$109.22$ &$8.90$ &$10.84$&$10.10$&$11.60$&$11.54$&$8.99$&$8.73$&$11.1$&10.95 &$8.18$   \\
$^{58}$Ni+$^{74}$Ge&$111.04$ &$7.00$ &$10.60$&$9.87$&$11.40$&$11.35$&$8.80$&$10.93$&$10.93 $&10.73 &$7.93$   \\
$^{40}$Ca+$^{124}$Sn&$118.95$ &$9.60$ &$10.96$&$10.17$&$11.80$&$11.72$&$9.23$&$8.95$&$11.35$&11.10 &$8.27$   \\
$^{28}$Si+$^{198}$Pt&$123.18$ &$9.80$ &$11.50$&$10.64$&$12.30$&$12.21$&$9.79$&$9.49$&$11.96 $& 11.53 &$9.56$   \\
$^{34}$S+$^{168}$Er&$124.24$ &$10.30$ &$11.35$&$10.53$&$12.20$&$12.09$&$9.68$&$9.38$&$11.81$& 11.49 &$8.78$   \\
$^{40}$Ar+$^{148}$Sm&$128.14$ &$8.30$ &$11.25$&$10.44$&$12.10$&$12.04$&$9.65$&$9.34$&$11.72 $&11.41 &$8.85$   \\
 $^{40}$Ar+$^{144}$Sm&$128.85$ &$8.30$ &$11.19$&$10.37$&$12.00$&$11.99$&$9.60$&$9.30$&$11.65 $&11.35 &$8.81$   \\
 $^{40}$Ca+$^{192}$Os&$165.42$ &$10.70$ &$11.54$&$10.62$&$12.20$&$12.55$&$10.35$&$10.04$&$12.21$&11.79 &$10.38$   \\
$^{40}$Ca+$^{194}$Pt&$169.40$ &$9.60$ &$11.53$&$10.60$&$-$&$12.97$&$10.40$&$10.09$&$12.22$&11.79 &$10.06$   \\
\hline
\hline
\label{t4}
\end{longtable}
\end{center}
\newpage
\begin{table}
\caption{Comparison of mean squared error between the experimental value and a model prediction for the quantities  B$_{fus}$ and R$_b$. }
\begin{tabular}{|c|c|c|c|c|c|c|c|c|c|c|}
\hline
 Quantity  & SEDF & Bass  & SW   & CW   & BW   & AW   & SPP   & WJS  &MR& \begin{tabular}[c]{@{}c@{}}Present\\ \end{tabular} \\ \hline
B$_{fus}$ & 7.63 & 15.14 & 0.82 & 0.84 & 1.61 & 2.41 & 11.41 & 3.76 & 6.88 &0.81\\
R$_b$     & 3.05 & 1.75  & 5.61 & 5.21 & 2.10 & 2.62 & 3.92  & -    & 4.42 & 0.80\\ \hline
\end{tabular}
\label{t5}
\end{table}
\newpage
\begin{table}
\caption{Comparison of experimental fusion barrier (B$_{fu}$) with the present formula for different reactions in the superheavy region. The last four reactions are used for the reaction contact time experiments \cite{albers2020zeptosecond}, where the WJS formula \cite{swiatecki2005fusion} has been used. We show here how much these values differ from the present formula.}

\begin{tabular}{ p{3cm}p{2cm}p{3cm}p{2cm}p{2cm}p{2cm}p{2cm}  }
\hline\hline
system             & z      & Expt.& Present&WDM\cite{myers2000nucleus} & WJS\cite{swiatecki2005fusion} & HCM\cite{manjunatha2018pocket}  \\ \hline
$^{32}S+^{232}Th$ & 154.52 & 155.73\cite{dutt2010systematic}  &   157.53&153.24&155.99&158.04\\
$^{48}Ca+^{208}Pb$ &  171.56& 173.40$\pm$ 0.1\cite{banerjee2019mechanisms}  &  173.11& 170.19&175.05&176.28\\
$^{50}Ti+^{208}Pb$ & 187.00 & 192.60$\pm$ 0.1\cite{banerjee2019mechanisms}  &191.82&187.62&192.42&194.04   \\
$^{48}Ti+^{208}Pb$ & 188.71 & 190.10\cite{dutt2010systematic}  &  193.02&188.18&194.35&195.10\\
$^{54}Cr+^{208}Pb$ &  202.79& 207.30$\pm$ 0.3\cite{banerjee2019mechanisms}  & 207.89&203.61& 210.22&210.68\\
$^{56}Fe+^{208}Pb$ & 218.64 & 223.00\cite{dutt2010systematic} &   221.73&220.71&228.14&228.67 \\
$^{64}Ni+^{208}Pb$ & 231.33 & 236.00\cite{dutt2010systematic} &    233.74&233.58&242.43&243.25\\
$^{70}Zn+^{208}Pb$ & 244.86 & 250.00\cite{dutt2010systematic} &   252.41&248.55&257.61&259.02\\
$^{86}Kr+^{208}Pb$ & 285.52 & 299.00\cite{dutt2010systematic} &   299.58&292.12&302.44&308.39\\
$^{86}Kr+^{208}Pb$ & 285.52 & 303.3\cite{ntshangase2007barrier} &   299.58&292.12&302.44&308.52\\\hline\hline
$^{50}Ti+^{249}Cf$&216.14&-&219.62&218.31&225.32&226.41\\
$^{54}Cr+^{248}Cm$&228.97&-&231.23&232.58&239.78&241.05\\
$^{58}Fe+^{244}Pu$&241.51&-&247.01&245.75&253.85&255.53\\
$^{64}Ni+^{238}U$&252.62&-&266.84&258.22&266.25&268.54\\
\hline\hline
\end{tabular}
\label{t6}
\end{table}
\begin{table}
\caption{Comparison of the fusion barrier heights (second column) ($V_B^{
exp}$) (unit is MeV) with various theoretical results calculated by the DFM $V^{DFM}_B$ (third column),
the DFM potential using the M3YReid interaction with zero-range exchange part ($V_B^{\delta R}$)
(fourth column) and using the Paris CDM3Y3 interaction with finite-range exchange part ($V_B^{Pf}$) (fifth column), respectively. All these DFM values have been taken from Table 3 of \cite{ganiev2020comparative}. In last column and last row, we give the present values ($V_B^{Pres.}$) and mean squared error of each set of theoretical values, respectively.  }
\begin{tabular}{|l|c|c|c|c|c|c|}
\hline
\multicolumn{1}{|c|}{Proj.+Targ.} & z & V$_B^{exp.}$ & V$_B^{DFM}$ & V$_B^{R\delta}$ & V$_B^{Pf}$ & V$_B^{Pres.}$ \\ \hline
$^{16}O+^{92}Zr$\cite{trotta2001sub} & 45.49 & 42.00 & 43.50 & 42.26 & 41.14 & 42.08 \\ \hline
$^{12}C+^{204}Pb$\cite{klaassen2001arctic} & 60.17 & 57.60 & 57.90 & 58.61 & 57.30 & 56.70 \\ \hline
$^{16}O+^{148}Sm$\cite{leigh1995barrier} & 63.51 & 59.80 & 61.10 & 61.19 & 59.61 & 60.03 \\ \hline
$^{17}O+^{144}Sm$\cite{leigh1995barrier} & 63.49 & 60.60 & 60.80 & 61.10 & 59.53 & 60.00 \\ \hline
$^{16}O+^{144}Sm$\cite{leigh1995barrier} & 63.91 & 61.00 & 61.40 & 61.53 & 59.94 & 60.42 \\ \hline
$^{28}Si+^{92}Zr$\cite{klaassen2001arctic} & 74.16 & 70.90 & 71.80 & 71.46 & 69.59 & 70.65 \\ \hline
$^{16}O+^{208}Pb$\cite{morton1999coupled} & 77.68 & 74.55 & 75.30 & 77.08 & 75.40 & 74.15 \\ \hline
$^{36}S+^{96}Zr$\cite{stefanini2000near} & 81.21 & 76.70 & 77.80 & 77.65 & 75.45 & 77.67 \\ \hline
$^{34}S+^{89}Y$\cite{mukherjee2002dominance} & 80.99 & 76.90 & 78.50 & 77.55 & 75.42 & 77.45 \\ \hline
$^{32}S+^{89}Y$\cite{mukherjee2002dominance} & 81.68 & 77.80 & 79.50 & 78.21 & 76.04 & 78.14 \\ \hline
$^{36}S+^{90}Zr$\cite{stefanini2000near} & 82.23 & 78.00 & 79.10 & 78.26 & 76.01 & 78.68 \\ \hline
$^{19}F+^{197}Au$\cite{leigh1995barrier} & 83.77 & 80.80 & 81.10 & 83.83 & 81.90 & 80.22 \\ \hline
$^{35}Cl+^{92}Zr$\cite{klaassen2001arctic} & 87.34 & 82.90 & 84.50 & 83.57 & 81.15 & 83.77 \\ \hline
$^{19}F+^{208}Pb$ \cite{hinde1999limiting}& 85.88 & 83.00 & 83.50 & 85.26 & 83.25 & 82.31 \\ \hline
$^{40}Ca+^{96}Zr$\cite{timmers1998case} & 100.01 & 94.60 & 96.50 & 97.01 & 94.32 & 96.41 \\ \hline
$^{40}Ca+^{90}Zr$\cite{timmers1998case} & 101.25 & 96.90 & 98.10 & 97.78 & 95.01 & 97.65 \\ \hline
$^{28}Si+^{144}Sm$\cite{dasgupta1994proceedings} & 104.86 & 104.00 & 102.20 & 104.37 & 101.72 & 101.28 \\ \hline
$^{40}Ca+^{124}Sn$\cite{scarlassara2000fusion} & 118.95 & 113.10 & 114.90 & 117.89 & 114.96 & 115.73 \\ \hline
$^{28}Si+^{208}Pb$\cite{hinde1995competition} & 128.10 & 128.10 & 126.10 & 130.90 & 128.08 & 126.34 \\ \hline
\multicolumn{3}{|c|}{Mean squared error} & 1.88 & 3.77 & 1.94 & 1.56 \\ \hline
\end{tabular}
\label{comparison-DFM}
\end{table}
\begin{table}
\caption{Comparison of the Woods-Saxon potential parameters $V_0$ (MeV), $r_0$ (fm), $a_0$ (fm) and $\hbar \omega$ from the present empirical formula and other models. The value marked with a superscript ($^*$) indicates Akyuz Winther (AkW) parameters \cite{akyuzparameterization} are altered to have agreement with measurements. Modified Akyuz Winther (MAkW) parameters are obtained by fixing the $V_0$ = 100 MeV and then calculating the values of $a_0$ and $r_0$ from the AkW model. The width $\hbar \omega$ is obtained from equation (48).}
\begin{center}
\centering
\begin{tabular}{|c|c|c|c|c|c|c|c|c|c|c|}
\hline
\multirow{2}{*}{Systems}  & \multicolumn{4}{c|}{Present work} &\multicolumn{6}{c|}{Other Models}\\ 
\cline{2-11}
&$V_0$&$a_0$&$r_0$& $\hbar\omega$&$V_0$&$a_0$&$r_0$&$\hbar\omega$&Model & Ref.\\ \hline 
$^{19}$F+$^{181}$Ta & 45.30&1.17 &1.15& 3.79 & 104.5 & 0.700 & 1.120 & 5.21&AkW$^*$ & \cite{shaikh2018investigation}  \\
$^{16}$O+$^{154}$Sm & 26.40&0.92 &1.24 & 4.19 & 100.0 & 1.060 & 1.019 & 3.85&MAkW & \cite{newton2004systematic} \\
$^{19}$F+$^{208}$Pb & 67.73&1.45 &1.06 & 3.51 & 100.0 & 1.060 & 1.059 & 4.34&MAkW & \cite{newton2004systematic}\\
$^{64}$Ni+$^{64}$Ni & 135.07 &1.76 &0.85& 2.61 & 75.98 & 0.676 & 1.190 & 5.06&AkW & \cite{scobel1976fusion} \\
$^{36}$S+$^{90}$Zr & 53.72 &1.18 &1.12 & 3.27 & 100.0 & 0.970 & 1.070 & 3.74&MAkW & \cite{newton2004systematic}\\
\hline
\end{tabular}
\label{t7}
\end{center}
\end{table}
 \newpage
\begin{center}
\begin{table}
\caption{Comparison of the present fusion barrier $B_{fus}^{Pres.}=V_B^{Pres.}$ Energy (in MeV) and radii (in fm) of the Coulomb barrier extracted from the DD-TDHF method \cite{washiyama2008energy}. $V^{DD}_B$ (high $E_{c.m.}$) refers to the barrier deduced for $E_{c.m.} > V^{FD}_B$, while $V^{DD}_B$ (low $E_{c.m}$.) corresponds to the lowest Coulomb barrier deduced from TDHF using $E_{c.m.} < V^{FD}_B$ and available experimental values. The reactions are listed in order of the increasing value of the $z$ parameter. Mean squared error (MSE) are given at the bottom row. See text for details.}
\begin{tabular}{||p{2.0cm}|c|c|c|c|c|c|c|c|c|c||}\hline\hline
\multicolumn{1}{||c|}{Reaction} &
  V$_B^{FD}$ &
  \begin{tabular}[c]{@{}c@{}}V$_B^{DD}$\\ (high E$_{c.m}$)\end{tabular} &
  \begin{tabular}[c]{@{}c@{}}V$_B^{DD}$\\ (low E$_{c.m}$)\end{tabular} &
  V$_B^{Pres.}$ &
  V$_B^{exp}$ &
  R$_B^{FD}$ &
  \begin{tabular}[c]{@{}c@{}}R$_B^{DD}$\\ (high E$_{c.m}$)\end{tabular} &
  \begin{tabular}[c]{@{}c@{}}R$_B^{DD}$\\ (low E$_{c.m}$)\end{tabular} &
  R$_B^{Pres.}$ &
  R$_B^{exp}$ \\\hline
$^{16}O+^{16}O$   & 10.2 & 10.13 & 10.12 & 10.41 & 10.61\cite{vaz1981fusion} & 8.4  & 8.46  & 8.52  &  & 7.91\cite{vaz1981fusion}  \\
$^{16}O+^{40}Ca$  & 23.5 & 23.36 & 23.07 & 23.91 & 23.06\cite{vaz1981fusion} & 9.2  & 9.18  & 9.50  &  & 9.21\cite{vaz1981fusion}  \\
$^{16}O+^{48}Ca$  & 23.0 & 22.77 & 22.48 & 23.01 &       & 9.4  & 9.50  & 9.75  &  &       \\
$^{40}Ca+^{40}Ca$ & 54.7 & 54.54 & 53.35 & 55.00 & 52.80\cite{dasgupta1998measuring}  & 9.8  & 9.82  & 10.32 & 8.72 &       \\
$^{40}Ca+^{48}Ca$ & 53.4 & 53.24 & 52.13 & 53.53 & 52.00\cite{newton2004systematics} & 10.1 & 10.09 & 10.56 & 8.88 & 9.99\cite{newton2004systematics}  \\
$^{48}Ca+^{48}Ca$ & 52.4 & 52.13 & 50.97 & 51.56 & 51.49\cite{newton2004systematics} & 10.3 & 10.38 & 10.82 & 9.02 & 10.16\cite{newton2004systematics} \\
$^{16}O+^{208}Pb$ & 76.0 & 75.91 & 74.51 & 74.15 & 74.52\cite{newton2004systematics} & 11.8 & 11.74 & 12.14 & 10.41 & 11.31\cite{newton2004systematics} \\
$^{48}Ca+^{90}Zr$ & 99.8 & 99.98 & 97.71 & 97.65 & 96.88\cite{newton2004systematics} & 10.8 & 10.63 & 11.27 & 8.55 & 10.53\cite{newton2004systematics}\\\hline
\begin{tabular}[c]{@{}c@{}}MSE\end{tabular} &3.51 &3.37 &0.34&1.74&-&0.17 & 0.13 &0.99 &3.63 &- \\\hline\hline
\end{tabular}
\label{t8}
\end{table}
\end{center}
\newpage
\begin{center}
\begin{longtable}{|p{2.0cm}| p{1.20cm} |p{1.20cm}|p{1.2cm}|p{1.3cm}|p{2.0
cm}| p{1.20cm}| p{1.20cm}| p{1.2cm}| p{1.5cm}|}
  \caption{Comparison of the interaction barrier $B_{int}$ for different systems between the present empirical model, Bass interaction model \cite{bass1974fusion} and available experimental values. The reactions are listed in order of the increasing value of the $z$ parameter. {Mean squared errors for Bass model and present formula are 15.22 and 1.74, respectively.}}\\
   \hline
System & $z$ &$Bass$ &$Pres.$ &$Exptl.$ &System & $z$ &$Bass$ &$Pres.$ &$Exptl.$\\  \hline
$^{32}$S+$^{24}$Mg&$31.69$ &$25.72$ & $15.56$&&$^{16}$O+$^{150}$Nd&$61.29$ &$57.41$ & $57.90$&\\
$^{32}$S+$^{27}$Al &$33.69$ &$27.71$ & $19.06$&&$^{16}$O+$^{148}$Nd&$61.46$ &$57.57$ & $58.13$&\\
$^{18}$O+$^{64}$Ni&$33.83$ &$28.37$ & $19.31$&&$^{12}$C+$^{238}$U&$65.04$&$62.63$&$62.38$&$62.2$\cite{viola1962total}\\
$^{18}$O+$^{62}$Ni&$34.05$ &$28.55$ & $19.68$&&$^{35}$Cl+$^{64}$Ni&$65.46$ &$60.18$ & $62.87$&\\
$^{18}$O+$^{60}$Ni&$34.27$ &$28.72$ & $20.06$&&$^{35}$Cl+$^{62}$Ni&$65.85$ &$60.49$ & $63.31$&$63.30$\cite{scobel1976fusion}\\
$^{16}$O+$^{64}$Ni&$34.36$ &$28.85$ & $20.20$&&$^{35}$Cl+$^{60}$Ni&$66.24$ &$60.81$ & $63.77$&\\
$^{18}$O+$^{58}$Ni&$34.51$ &$28.91$ & $20.45$&&$^{35}$Cl+$^{58}$Ni&$66.65$ &$61.13$ & $64.24$&$64.2$\cite{scobel1976fusion}\\
$^{16}$O+$^{62}$Ni&$34.58$ &$29.03$ & $20.58$&&$^{14}$N+$^{238}$U&$74.82$&$72.83$&$73.47$&$73.4$\cite{viola1962total}\\
$^{16}$O+$^{60}$Ni&$34.81$ &$29.21$ & $20.97$&&$^{16}$O+$^{205}$Tl&$76.99$&$74.60$&$75.85$&$77.0$\cite{le1972nuclear}\\
$^{18}$O+$^{65}$Cu&$34.93$ &$29.47$ & $21.17$&&$^{16}$O+$^{238}$U&$84.43$&$82.92$&$83.84$&$82.5$\cite{viola1962total}\\
$^{18}$O+$^{64}$Cu&$35.05$ &$29.39$ & $21.35$&&$^{35}$Cl+$^{90}$Zr&$87.71$ &$83.57$ & $87.29$&\\
$^{18}$O+$^{63}$Cu&$35.15$ &$29.64$ & $21.54$&&$^{40}$Ar+$^{110}$Pd&$100.84$ &$98.01$ & $100.77$&\\
$^{35}$Cl+$^{27}$Al&$35.24$ &$29.28$ & $21.69$&&$^{35}$Cl+$^{124}$Sn&$102.93$ &$100.38$ & $102.89$&\\
$^{16}$O+$^{65}$Cu&$35.47$ &$29.95$ & $22.07$&&$^{20}$Ne+$^{238}$U&$103.24$&$102.81$&$103.19$&$102.0$\cite{viola1962total}\\
$^{18}$O+$^{70}$Zn&$35.59$ &$30.21$ & $22.28$&&$^{35}$Cl+$^{116}$Sn&$104.31$ &$101.52$ & $104.28$&\\
$^{16}$O+$^{63}$Cu&$35.69$ &$30.13$ & $22.44$&&$^{40}$Ar+$^{164}$Dy&$133.57$&$133.93$&$133.37$&$135.0$\cite{le1971behavior}\\
$^{18}$O+$^{68}$Zn&$35.81$ &$30.30$ & $22.63$&&$^{40}$Ar+$^{197}$Au&$153.92$ &$156.45$ & $153.64$&\\
$^{18}$O+$^{66}$Zn&$36.02$ &$30.55$ & $22.98$&&$^{40}$Ar+$^{208}$Pb&$157.95$ &$161.08$ & $157.69$&\\
$^{16}$O+$^{70}$Zn&$36.14$ &$30.71$ & $23.17$&&$^{40}$Ar+$^{238}$U&$172.19$&$177.18$&$172.17$&$171.0$\cite{bass1973threshold}\\
$^{18}$O+$^{64}$Zn&$36.25$ &$30.73$ & $23.35$&&$^{48}$Ti+$^{208}$Pb&$188.72$&$194.41$&$189.32$&$190.1$\cite{mitsuoka2007barrier}\\
$^{16}$O+$^{68}$Zn&$36.36$ &$30.89$ & $23.52$&&$^{54}$Cr+$^{208}$Pb&$202.79$&$209.91$&$204.27$&$205.8$\cite{mitsuoka2007barrier}\\
$^{16}$O+$^{66}$Zn&$36.58$ &$31.05$ & $23.89$&&$^{54}$Cr+$^{207}$Pb&$202.99$ &$210.07$ & $204.48$&\\
$^{16}$O+$^{64}$Zn&$36.81$ &$31.23$ & $24.26$&&$^{52}$Cr+$^{208}$Pb&$203.78$ &$210.79$ & $205.34$&\\
$^{12}$C+$^{152}$Sm&$48.78$ &$44.68$ & $41.95$&&$^{56}$Fe+$^{208}$Pb&$218.64$&$227.02$&$221.55$&$223.0$\cite{mitsuoka2007barrier}\\
$^{35}$Cl+$^{48}$Ti&$54.16$ &$48.39$ & $49.06$&&$^{58}$Ni+$^{208}$Pb&$234.38$&$244.03$&$239.19$&$236.0$\cite{mitsuoka2007barrier}\\
$^{16}$O+$^{134}$Ba&$58.66$ &$54.44$ & $54.72$&&$^{70}$Zn+$^{208}$Pb&$244.87$&$256.33$&$251.22$&$250.6$\cite{mitsuoka2007barrier}\\
$^{12}$C+$^{205}$Tl&$59.37$&$56.29$&$55.59$&$56.0$\cite{le1972nuclear}&$^{84}$Kr+$^{232}$Th&$307.86$&$326.78$&$328.62$&$332.0$\cite{bass1973threshold}\\
$^{12}$C+$^{209}$Bi&$60.55$&$57.57$&$57.04$&$57.0$\cite{le1972nuclear}&$^{84}$Kr+$^{238}$U&$313.14$&$332.81$&$-$&$333.0$\cite{moretto1972shell}\\
\hline
\label{t9}
 \end{longtable}
 \end{center}
\newpage
\newpage
\newpage
\newpage
\end{document}